\documentclass[11pt]{article}

\usepackage{calc}
\usepackage{xspace}
\usepackage{subfig}
\usepackage{amsmath}
\usepackage{amssymb}
\usepackage{amsfonts}
\usepackage{mathrsfs}
\usepackage{enumerate}
\makeatletter
\@addtoreset{equation}{section}

\makeatother
\usepackage{feynmf}
\usepackage{graphicx}
\usepackage{amssymb,amsfonts,amsmath}
\usepackage{accents}

\def\beq{\begin{equation}}
\def\eeq{\end{equation}}
\def\bea{\begin{eqnarray}}
\def\eea{\end{eqnarray}}

\begin{document}

\renewcommand{\thefootnote}{\fnsymbol{footnote}}
\newcommand{\inst}[1]{\mbox{$^{\text{\textnormal{#1}}}$}}
\begin{flushright}
July, 2020
\end{flushright}
\begin{center}
{\LARGE Simplicial Gravity with Coordinates}\\[4ex]
%
{\large
Alessandro D'Adda\footnote{\texttt{dadda@to.infn.it}}}

%
{\large\itshape
 INFN Sezione di Torino, and\\
Arnold-Regge Center,\\
via P. Giuria 1, 10125 Torino, Italy}

\end{center}
\bigskip
\setcounter{footnote}{0}
\renewcommand{\thefootnote}{\arabic{footnote}}

\begin{abstract}

We present a formulation of Regge Calculus where arbitrary coordinates are associated to each vertex of the simplicial complex and the fundamental degrees of freedom are given by the metric  $g_{\mu\nu}(\alpha)$ on each simplex $\alpha$.

The lengths of the edges, which are the usual degrees of freedom of Regge Calculus, are thus determined and are left invariant under arbitrary transformations of the discrete set of coordinates, provided the metric transforms accordingly.

Invariance under coordinate transformations entails tensor calculus and our formulation then follows closely the usual formalism of the continuum theory. This includes a definition of partial derivative which stems from a generalization to  simplicial lattices of the symmetric finite difference operator on a cubic lattice.

The definitions of parallel transport, Christoffel symbol, covariant derivatives and Riemann curvature tensor  follow in a rather natural way establishing a kind of dictionary between continuum and simplicial lattice quantities. In this correspondence Einstein action becomes Regge action with the deficit angle $\theta$ replaced by $\sin \theta$.

The correspondence with the continuum theory can be extended to actions with higher powers of the curvature tensor, to the vielbein formalism and to the coupling of gravity  with matter fields (scalars, fermionic fields including spin $3/2$ fields and gauge fields) which are then determined unambiguously and discussed in the paper.

An action on the simplicial lattice for $N=1$ supergravity in $4$ dimensions is derived in this context.

Another relavant result is that Yang-mills actions on a simplicial lattice consist, even in absence of gravity, of two plaquettes terms, unlike the one plaquette Wilson action on the hypercubic lattice.

 An attempt is also made to formulate a discrete differential calculus to include  differential forms of higher order and the gauging of free differential algebras in this scheme.  However this leads to form products that do not satisfy associativity and distributive law with respect to the $d$ operator. 

A proper formulation of theories that contain higher order differential forms in the context of Regge Calculus is then still lacking.

\end{abstract}

PACS codes: 04.60.Nc, 11.15.Ha, 11.10.−z

Keywords: Regge Calculus, Descrete Gravity, Lattice gauge Theory

\newpage
\section{Introduction}
\label{0.1}

The title of Regge's seminal paper of 1961 \cite{Regge:1961px} ``General relativity without coordinates'' emphasizes a crucial aspect of his approach to discrete gravity, namely that it does away with the notion of coordinates and formulates general relativity purely in terms of geometrical quantities: lengths, volumes, angles, etc.

This was in itself an extraordinary achievement. 
In the continuum theory absolute differential calculus, or tensor calculus, plays a fundamental role in the mathematical formulation of general relativity.  Invariance under general coordinate transformations follows directly from the  principle of equivalence in its most general form, namely that all reference frames are equivalent in the description of the physical world and that the only real observables are the underlying geometric properties of space-time, which are the building blocks of Regge's formulation. 
 
The basic ideas of Regge Calculus are well known: the smooth $d$-dimensional space-time manifold of the continuum formulation is replaced by a triangulated manifold made of piecewise flat $d$-dimensional simplices glued together by identifying in pairs their $d-1$ dimensional faces  . The geometrical properties of this manifold are  determined by   the  lengths of all its one dimensional edges: in fact each $d$-dimensional simplex is completely fixed by the lengths of its $\frac{d(d+1)}{2}$ edges.

The curvature is associated to the $d-2$ dimensional subspaces, the hinges, and is given for each hinge $h$ by the deficit angle $\theta_h$ defined as $2 \pi$ minus the sum of the dihedral angles between the faces of the simplices which the hinge $h$ belongs to.
In a flat space $\theta_h$ is zero for all $h$, as clearly shown by the two dimensional case, where the hinges are points (dimension zero) and $\theta_h$ the complement to $2 \pi$ of the sum of the angles meeting at that point.

The discrete version of  Einstein action is then given by:
\begin{equation}
S_R = \sum_h |V|_h \theta_h       \label{reggeaction}
\end{equation}
where $|V|_h$ is the volume of the hinge $h$.

Following Regge's original paper a great number of different formulations and  approaches to Regge calculus appeared. We shall not even try to go over the huge literature on the subject, which can be found in the review paper of ref. \cite{Regge:2000wu}, recently updated in ref.\cite{Barrett:2018ybl}

Some of the new proposals mantained the same purely geometrical approach of the original Regge paper, like the so called Area Regge Calculus \cite{Barrett:1997tx} where in four dimensions the areas of the triangles are chosen as fundamental degrees of freedom in place of  the edges' lengths.

Coupling gravity with matter fields, and in particular with fermions, requires however the introduction of vielbeins, and hence of some kind of local coordinates, on the simplicial complex . 
 
This was done in ref.\cite{Caselle:1989hd} and \cite{Kawamoto:1990hk}, where a euclidean reference frame is introduced in each simplex, and the  degrees of freedom are defined on the links of the dual lattice as the Poincar\'e transformations needed to rotate the reference frame defined on a simplex $\alpha$ into the one of a contiguous simplex $\beta$. The action is the one of a gauge theory on the dual lattice with a local Poincar\'e invariance, but it is eventually equivalent to Regge action of eq.(\ref{reggeaction}) although with the deficit angle $\theta_h$ replaced by $\sin ( \theta_h)$.

 A first order formalism is possible in this framework, and the presence of a local Lorentz group makes the coupling of fermions to gravity possible.

In this paper, while keeping the original Regge's triangulation of space-time, we reintroduce space time coordinates trying to keep the formalism as close as possible to the continuum formulation. This is done  by associating arbitrary space time coordinates to each vertex of the simplicial complex and a costant metric tensor $g_{\mu\nu}(\alpha)$ to each simplex $\alpha$\footnote{A similar parametrization has been used by Khatsymovsky in ref.\cite{Khatsymovsky:2015dqa}.}. 
The length of all the edges, which are the degrees of freedom of Regge Calculus, are then entirely fixed and are preserved by arbitrary transformations of the coordinates provided the metric tensor in each simplex is transformed accordingly (Section \ref{s2}).

With this choice of degrees of freedom  gravity can be formulated on a simplicial complex following step by step the classical textbook formalism of continuum general relativity. This includes  a discretized version of tensor calculus which can be formalized to assure invariance under coordinate transformations (Section \ref{sectens}).  Another fundamental step is the definition of partial derivative on the simplicial lattice, which generalizes in a non trivial way the symmetric finite difference operator on the hypercubic lattice (Section \ref{derivativessl}). Parallel transport can then be defined to make derivatives covariant (Section \ref{PT})  and eventually the analogue of the Riemann curvature tensor is obtained (Section   \ref{Riemannsect}). 

As a result a kind of dictionary is established that allows to translate any gravitational action in the continuum into a corresponding action on the simplicial complex. Within this correspondence the Einstein action is naturally translated into Regge's action but  with the deficit angle $\theta$ replaced by $\sin\theta$ as in ref.\cite{Caselle:1989hd}(Section   \ref{Riemannsect}).

Gravitational actions with higher derivatives terms and Brans-Dicke type of actions can also be included in this scheme, and a definite prescription for their formulation on a simplicial lattice is given in Section \ref{HDBD}.

Brans-Dicke action involves the coupling of the gravitational field to scalar fields. The coupling of gravity to matter fields with higher spin, such as gauge fields and fermions is the subject of the last sections of the paper. Gauge fields are defined on the links of the dual lattice and the field strength on the dual lattice plaquettes (i.e. the hinges of the simplicial lattice) whose number of sides in not fixed. Yang-Mills action is obtained by coupling with the metric tensor two plaquettes that have a site (that is a simplex of the original lattice) in common and  is therefore rather different, even in absence of gravity, from the one plaquette term of the standard formulation on an hypercubic lattice
 (Section \ref{sectiongauge}).

The vielbain formalism is introduced in Section \ref{vielbainsect} by following the same approach used in ref.\cite{Caselle:1989hd}, that is by introducing in each simplex $\alpha$ a euclidean reference frame defined up to an arbitrary Lorentz rotation. The vielbeins in $\alpha$ are identified with the components of the local coordinate transformation from the general frame originally defined in $\alpha$ by the coordinate choice to the local euclidean frame. As in the continuum theory the vielbein transform under both the general coordinate transformation and the local Lorentz transformations, which constitute a local symmetry group of the theory and can be treated according to the scheme already outlined in section \ref{sectiongauge}. As in ref.\cite{Caselle:1989hd} the Lorentz connections are defined on the links of the dual lattice and are the gauge fields associated to the local Lorentz rotations.

The introduction of the vielbains and of the local Lorentz group makes  the coupling of fermionic fields to gravity on a simplicial lattice possible exactly as in the continuum case. This is discussed in Section \ref{fermionsect}. Having established a discrete version of tensor calculus  this coupling can be easily extended to fermionic fields that transform as vectors under general coordinate transformations, like for instance the gravitino. It is then possible to write a discrete action that corresponds to $D=4$ and $N=1$ supergravity in the continuum.

In order to have a complete correspondence between continuum and simplicial lattice  theories one should include one more set of fields, namely the $p$-form potentials (with $p >1$) that arise from the gauging of free differential algebras. These fields play an important role for instance in higher dimensional supergravity theories. 
In Section \ref{pformsection}. we discuss this point and find that a straightforward extension to these fields of the correspondence established in the previous sections leads to field strengths ($p+1$ forms in the continuum) that are not gauge invariant.

This is probably related to the fact that, in spite of the invariance under coordinates transformations (which however involves only a descrete set of points), our formulation is equivalent to Regge Calculus and does not have invariance under diffeomorphisms.
A consistent formulation of differential forms on a simplicial lattice\footnote{Actually the precise correspondence would be with the lattice dual to the original simplicial lattice.} would probably be the answer to the problem of including $p$-form potential. Although this was  beyond the original purpose of the paper an attempt was made in this direction leading to a definition of discrete differential forms that, although elegant, has a non-associative product. More seriously, the product does not obey the usual distribution law with respect to the $d$ operator. This is discussed for completeness in the Appendix.

 \section{Simplicial Gravity with Coordinates}
\label{s2}
Let $\alpha$ be a $d$-dimensional simplex and $i,j,\dots$ labels for its $d+1$ vertices.
In Regge calculus the simplex is completely identified by giving the $\frac{d(d+1)}{2}$ lengths $l_{ij}$ of the edges joining the vertices $i$ and $j$. The lengths $l_{ij}$ have to satisfy triangular inequalities, but are otherwise arbitrary. They constitute the fundamental degrees of freedom of Regge's discrete gravity.

There are alternative ways to identify the simplex $\alpha$. One is to associate a coordinate $x^\mu_i$ ($\mu = 1,2,\dots d$) to each vertex $i$ {\it and} a constant  metric $g_{\mu\nu}(\alpha)$, in general not euclidean, to each simplex $\alpha$.
The lengths $l_{ij}$ of the edges are then determined and given by:
\begin{equation}
l_{ij}^2 = g_{\mu\nu}(\alpha) \left(x_i^\mu - x_j^\mu\right)\left(x_i^\nu-x_j^\nu\right) \label{lengths}.
\end{equation}
Conversely, if the lengths $l_{ij}$ are given and the coordinates $x^\mu_i$ of the vertices are chosen in an arbitrary way, then eq.s (\ref{lengths}) provide a set of $d(d+1)/2$ equations in the $d(d+1)/2$ unknown components of the metric  $g_{\mu\nu}(\alpha)$. These equations have a unique solution provided the determinant of the $\frac{d(d+1)}{2}\times\frac{d(d+1)}{2}$ matrix $\Delta_{ij,\mu\nu}$  of their coefficients is not vanishing:
\begin{equation}
\det \Delta_{ij,\mu\nu} \equiv  \det \left\{ \left(x_i^\mu - x_j^\mu\right)\left(x_i^\nu-x_j^\nu\right)\right\}\neq 0.
\label{bigdet} \end{equation}
The determinant of $\Delta_{ij,\mu\nu}$ can be calculated and it is given by:
\begin{equation}
\det \Delta_{ij,\mu\nu} = \left[ \det \left(x^\mu_i-x^\mu_{d+1}\right) \right]^{d+1},~~~~~~~~~~~~i=1,\dots,d \label{bigdet2} \end{equation}
so that the  condition (\ref{bigdet}) is satisfied iff the determinant of the differences $x^\mu_i-x^\mu_{d+1}$ is different from zero:
\begin{equation}
\det \left(x^\mu_i-x^\mu_{d+1}\right) \neq 0~~~~~~~~~~~~i=1,\dots,d  .     \label{det}
\end{equation}
The last condition insures that the simplex $\alpha$ is not degenerate in $d$ dimensions.
So a simplicial manifold can be characterized by assigning, instead of the edges' lengths as in  Regge Calculus, the coordinates of all the vertices and the metric of each simplex. This is essentially the same as in the usual formulation of Einstein's gravity in the continuum.  As in the continuum the choice of the coordinates is arbitrary, provided for each simplex the determinant condition (\ref{det}) is satisfied, and we expect the theory to be invariant under general coordinate transformations, that is to depend only on geometrical invariants such as $l_{ij}$.

Notice however that the components of the metric tensors belonging to different simplices are not all independent. Consider in fact two simplices $\alpha$ and $\beta$ which have in common a $d-p$ dimensional sub-simplex.  Their common edges have lengths $l_{ij}$ that cannot depend upon the fact of being considered as part of $\alpha$ or as part of $\beta$. 
Then from (\ref{lengths}) we have:
\begin{equation}
\left[g_{\mu\nu}(\alpha) - g_{\mu\nu}(\beta)\right] \left( x_i^\mu - x_j^\mu \right) \left( x_i^\nu - x_j^\nu \right) =0~~~~~~i,j \in \alpha\cap\beta.   \label{commonface} \end{equation}
 Eq.s (\ref{commonface}) should be regarded as constraints to be implemented (which may not be easy)  whenever the metric is varied or is integrated upon in the functional integral.
 As a result the number of degrees of freedom per simplex associated to the metric is much smaller than $\frac{d(d+1)}{2}$.
 In fact, consider a simplex $\alpha$ with a given metric $g_{\mu\nu}(\alpha)$,  and a simplex $\beta$ that has a $d-1$-dimensional face in common with $\alpha$. In this case $\alpha\cap\beta$ has $d$ vertices and $\frac{d(d-1)}{2}$ links $ij$; so it follows from  (\ref{commonface})  that  if   $g_{\mu\nu}(\alpha)$ is fixed only  $\frac{d(d+1)}{2}-\frac{d(d-1)}{2} = d$ components of $g_{\mu\nu}(\beta)$  can be chosen independently from $g_{\mu\nu}(\alpha)$.
 Furthermore, if a simplex $\gamma$ has a face in common with $\beta$ it still has a $(d-2)$-dimensional hinge, namely $\frac{(d-1)(d-2)}{2}$ links, in common with $\alpha$ and the componenents of $g_{\mu\nu}(\gamma)$ independent from $g_{\mu\nu}(\alpha)$ are just $2d-1$. It is easy to conclude that two simplices have completely independent metrics only if they are separated by at least $d$ simplices, namely in the dual\footnote{Here and in the following we define the dual lattice as the lattice obtained by a Voronoi tassellation of the simplicial complex. The vertices of the dual lattice are then the circumcenters $c(\alpha)$ of the simplices $\alpha$ and its links are the lines joining the circumcemters of neighbouring simplices. Although it is not strictly necessary we shall assume that the circumcenters are always inside the corresponding simplex, namely that the simplicial complex is a Delaunay triangulation.}  lattice if they are vertices separated by at least $d$ links.
 
 Consider now a simplicial submanifold $\mathcal{R}$ (we assume for simplicity that it has the topology of a sphere) made of a large number on simplices. As discussed before an additional simplex $\gamma$ increases the number of degrees of freedom by $d$ if $\gamma$ is attached to $\mathcal{R}$ by a single face, or  $0$ if $\gamma$ has two faces in common with $\mathcal{R}$\footnote{In that case $\gamma$ is the last simplex needed to complete the simplices insisting on a hinge.}.
 The latter case being only a (presumably small) fraction of the total we may conclude that the number of degrees of freedom per simplex associated to the metric is not $\frac{d(d+1)}{2}$, as one would naively expect from a correspondence with the continuum case, but is of order $d$.

  \section{General coordinates transformations and tensor calculus.}
\label{sectens}
Let us consider now a general coordinate transformation on the simplicial manifold:
\begin{equation}
x^\mu_i \Longrightarrow x'^\mu_i~~~~~~~~i=1,\dots,N   \label{gct} \end{equation}
where $N$ is now the total number of vertices in the manifold. The metric $g_{\mu\nu}(\alpha)$ of each simplex $\alpha$ should transform under (\ref{gct}) into a new metric $g'_{\mu\nu}(\alpha)$ in such a way to leave all the edges lengths $l_{ij}$ invariant.

Let now $x^\mu_i$ be the coordinates on the vertices of a specific simplex $\alpha$. Then the general coordinate transformation (\ref{gct}), restricted to the vertices  in $\alpha$ can always be written as:
\begin{equation}
x'^\mu_i = \Lambda_{\ \nu}^\mu(\alpha) x^\nu_i + \Lambda^\mu(\alpha)~~~~~~i\in \alpha \label{gctalpha} \end{equation}
which also implies:
\begin{equation}
x'^\mu_i-x'^\mu_j = \Lambda_{\ \nu}^\mu(\alpha) \left(x_i^\nu-x_j^\nu\right)~~~~~~i,j\in \alpha. \label{tp} \end{equation}
The matrix $ \Lambda_{\ \nu}^\mu(\alpha) $ is the discrete analogue of $\frac{\partial x'^\mu}{\partial x^\mu}$ and eq.(\ref{tp}) can be used to define the transformation properties of a contravariant vector under general coordinate transformation on the simplicial manifold:
\begin{equation}
A'^\mu(\alpha) = \Lambda^\mu_{\ \nu}(\alpha) A^\nu(\alpha).    \label{contratp}
\end{equation}
We shall assume that $\det \Lambda_{\ \nu}^\mu(\alpha) \neq 0 $ for all $\alpha$. In fact it is clear from (\ref{tp}) and (\ref{det}) that this is the necessary and sufficient condition for (\ref{det}) to be preserved under (\ref{gct}).
Notice that if $i$ and $j$ in (\ref{tp}) belong to both simplices $\alpha$ and $\beta$ then from  (\ref{tp}) we have:
\begin{equation}
\left[ \Lambda^\mu_{\ \nu}(\alpha) -  \Lambda^\mu_{\ \nu}(\beta) \right] \left( x_i^\nu-x_j^\nu \right) = 0~~~~~~~~~~~~~~~i,j \in \alpha\cap\beta.   \label{commonLa} \end{equation}
Eq. (\ref{commonLa}) follows automatically from the restriction of (\ref{gct}) to the simplices $\alpha$ and $\beta$. However an alternative way of defining a general coordinate transformation is to assign, in place of (\ref{gct}),  the matrices $ \Lambda^\mu_{\ \nu}(\alpha)$ and $\Lambda^\mu(\alpha)$ for each simplex $\alpha$. In that case eq.s (\ref{commonLa}) should be regarded as  constraints to be implemented on $ \Lambda^\mu_{\ \nu}(\alpha)$.

Tensor calculus can now be formulated on the lattice: tensors with covariant and contravariant indices can  be defined as quantities that transform with $\Lambda$ for each contravariant index and with $\Lambda^{-1}$ for each covariant index:
\begin{equation}
A'^{\mu_1\cdots\mu_h}_{\ \ \ \ \ \ \ \ \nu_1\cdots\nu_k}(\alpha) = \Lambda^{\mu_1}_{\ \rho_1}(\alpha)\cdots\Lambda^{\mu_h}_{\ \rho_h}(\alpha)(\Lambda^{-1})^{\sigma_1}_{\ \nu_1}(\alpha)\cdots(\Lambda^{-1})^{\sigma_k}_{\ \nu_k}(\alpha)~ A^{\rho_1\cdots\rho_h}_{\ \ \ \ \ \ \ \ \sigma_1\cdots\sigma_k}(\alpha). \label{tensor}
\end{equation}

The metric $g_{\mu\nu}(\alpha)$ transforms as a covariant tensor of rank $2$. In fact from (\ref{tp}) the requirement  that the lengths $l_{ij}$ given in (\ref{lengths}) are  invariant under general coordinate transformations gives:
\begin{equation}
g'_{\mu\nu}(\alpha) = (\Lambda^{-1})_{\ \mu}^{\rho}(\alpha)~ (\Lambda^{-1})_{\ \nu}^{\sigma}(\alpha)~ g_{\rho\sigma}(\alpha). \label{gtransform} \end{equation}

Quantities that are invariant under general coordinate transformations, and hence have an intrinsic geometrical meaning can now be constructed. The simplest is the volume  $V(\alpha)$ of the simplex, which is the discrete analogue of  the invariant volume $\sqrt{g}~ d^dx$ of the continuum theory,  and is given by:
\begin{equation}
V(\alpha) =\frac{1}{d!} \epsilon_{\mu_1\mu_2\cdots\mu_d}\left(x_1^{\mu_1}-x_{d+1}^{\mu_1}\right)\cdots\left(x_d^{\mu_d}-x_{d+1}^{\mu_d}\right) \sqrt{\det g_{\mu\nu}(\alpha)}. \label{simplexvolume} \end{equation}
Notice that the value of $V(\alpha)$ changes sign, due to the antisymmetry of the $\epsilon$ symbol, if an odd permutation of the vertices  is performed. We shall assume in the future that the order of the $x_i$'s in (\ref{simplexvolume}) is such that $V(\alpha)$ is positive\footnote{Alternatively the absolute value can be taken at the r.h.s. Notice also that the choice of the label $d+1$ for the reference vertex is irrelevant modulo a sign factor coming from the antisymmetric tensor.}. The invariance of $V(\alpha)$ under general coordinate transformations follows from  (\ref{tp}) and (\ref{gtransform}).

Given a simplex $\alpha$ there are $d+1$ neighbouring simplices that have a  $d-1$ dimensional face in common with $\alpha$. We shall denote such simplices as $\alpha_i$, where the index $i$ denotes the vertex of $\alpha$ which is not in $\alpha_i$: $x_i \notin \alpha_i$.

Let us denote by $\alpha\cap\alpha_i$ the $d-1$ dimensional face that $\alpha$ and $\alpha_i$ have in common.

We can define then the following covariant vector:
 
\begin{eqnarray}
&V&^{(\alpha\cap\alpha_i)}_{ \mu}(\alpha) = d\: \frac{\partial V(\alpha)}{\partial x_i^\mu} \nonumber \\& = &\frac{ \sqrt{\det g_{\mu\nu}(\alpha)} }{(d-1)!} \epsilon_{\nu_1\dots \mu \dots \nu_{d}} \left(x_1^{\nu_1}-x_{d+1}^{\nu_1}\right)\dots \langle i \rangle \dots \left(x_{d}^{\nu_{d}}-x_{d+1}^{\nu_{d}}\right)\label{area} \end{eqnarray}
where the index $\mu$ in the $\epsilon$ symbol is in the $i$-th position and the $\langle i \rangle$ bracket means that the term $\left(x_i^{\nu_i}-x_{d+1}^{\nu_i}\right)$ in the product is missing.

The vector $V^{(\alpha\cap\alpha_i)}_{ \mu}(\alpha)$ is orthogonal to the face $\alpha\cap\alpha_i$:
\begin{equation}
\left( x_r^\mu - x_s^\mu \right) V^{(\alpha\cap\alpha_i)}_{ \mu}(\alpha) = 0~~~~~~~~~r,s \neq i  \label{ortho} \end{equation}
and it can be shown to be pointing toward the outside of $\alpha$. 
Eq. (\ref{area}) also implies:
\begin{equation}
\left(  x_i^\mu - x_r^\mu \right) V^{(\alpha\cap\alpha_i)}_{ \mu}(\alpha) = d~ V(\alpha)~~~~~~~r \neq i.       \label{vv} \end{equation}

The modulus of $ V^{(\alpha\cap\alpha_i)}_{ \mu}(\alpha)$ is equal to the $d-1$ dimensional volume $V(\alpha\cap\alpha_i)$  of $\alpha\cap\alpha_i$, so that we can write \footnote{This follows from the observation that the modulus does not depend on the choice of the coordinates and it is easily seen by choosing the coordinates of vertices of $\alpha$ in such a way that $g_{\mu\nu}(\alpha)=\eta_{\mu\nu}$, where  $\eta_{\mu\nu}$ is the euclidean metric}:
\begin{equation}
 V^{(\alpha\cap\alpha_i)}_{ \mu}(\alpha) = V(\alpha\cap\alpha_i)~ n_\mu^{(\alpha\cap\alpha_i)}(\alpha)  \label{nmu} \end{equation}
 where $ n_\mu^{(\alpha\cap\alpha_i)}(\alpha)$ is a vector orthogonal to $ \alpha\cap\alpha_i$, pointing to the outside of $\alpha$ and with modulus $1$:
 \begin{equation}
  n_\mu^{(\alpha\cap\alpha_i)}(\alpha) g^{\mu\nu}(\alpha)  n_\nu^{(\alpha\cap\alpha_i)}(\alpha) = 1. \label{modulusn} \end{equation}
A unit vector $ n_\mu^{(\alpha\cap\alpha_i)}(\alpha_i)$ orthogonal to $\alpha\cap\alpha_i$ can be obtained starting from $\alpha_i$ instead of $\alpha$. 
It can be shown then from  (\ref{area}), (\ref{nmu}) and (\ref{simplexvolume}) that   $ n_\mu^{(\alpha\cap\alpha_i)}(\alpha_i)$ is proportional to  $ n_\mu^{(\alpha\cap\alpha_i)}(\alpha)$. More precisely we have:
 \begin{equation}
 \frac{n_\mu^{(\alpha\cap\alpha_i)}(\alpha)}{\sqrt{\det g_{\mu\nu}(\alpha)} } = -  \frac{n_\mu^{(\alpha\cap\alpha_i)}(\alpha_i)}{\sqrt{\det g_{\mu\nu}(\alpha_i)} }
 \label{nmuprop}
 \end{equation}
 where the minus sign is due to the orientation convention.
 
The simplex $\alpha$ and any two neighbouring simplices $\alpha_i$ and $\alpha_j$ have a $d-2$ dimensional simplex (hinge) $h_{ij}$ in common. The set of the $d-1$ vertices of $h_{ij}$ is the set of the vertices of $\alpha$ where the vertices labeled $i$ and $j$ have been removed. Since there are $\frac{d(d+1)}{2}$ ways of removing two vertices from $\alpha$ there are $\frac{d(d+1)}{2}$ distinct  hinges belonging to $\alpha$.  

In analogy with what we have done for the faces, we can associate to the hinge $h_{ij}$ the covariant tensor
\begin{eqnarray}
&V&^{(h_{ij})}_{ \mu_1\mu_2}(\alpha) = d (d-1) \frac{\partial^2 V(\alpha)}{\partial x_j^{\mu_2} \partial x_i^{\mu_1}} = \nonumber \\ &=& \frac{\sqrt{\det g_{\mu\nu}(\alpha)}}{(d-2)!} \epsilon_{\nu_1\dots\mu_1\dots \mu_2\dots \nu_{d}} \left(x_1^{\nu_1}-x_{d+1}^{\nu_1}\right)\cdots\langle  i j\rangle \cdots \left(x_{d}^{\nu_{d}}-x_{d+1}^{\nu_{d}}\right)  \label{hinge} \end{eqnarray}
where the indices $\mu_1$ and $\mu_2$ in the $\epsilon$ antisymmetric tensor are respectively in the $i$-th and the $j$-th position and the symbol $\langle ij \rangle$ means that the terms $(x_i^{\nu_i} -x_{d+1}^{\nu_{i}} ) $ and $(x_j^{\nu_j} -x_{d+1}^{\nu_{j}} ) $  are omitted in the product at the r.h.s. of (\ref{hinge}).

Notice that $V^{(h_{ij})}_{ \mu_1\mu_2}(\alpha)$ is not only antisymmetric in the tensor indices  $ \mu_1$ and $\mu_2$ but also under exchange of $i$ and $j$:
\begin{equation}
 V^{(h_{ij})}_{ \mu_1\mu_2}(\alpha) = - V^{(h_{ji})}_{ \mu_1\mu_2}(\alpha). \label{hingeor} \end{equation}
$V^{(h_{ij})}_{ \mu_1\mu_2}(\alpha)$ and $V^{(h_{ji})}_{ \mu_1\mu_2}(\alpha)$ correspond to the two different orientations of the hinge, which are better viewed by going to the dual lattice where the $d-2$ dimensional hinge corresponds to a 2-dimensional plaquette.

It can be easily seen from (\ref{hinge}) that $V^{(h_{ij})}_{ \mu_1\mu_2}(\alpha)$ is orthogonal to the hinge $ h_{ij}$:
\begin{equation}
\left( x_r^\mu - x_s^\mu \right) V^{(h_{ij})}_{ \mu\nu}(\alpha) = 0~~~~~~~~~\forall~ r,s \neq i,j . \label{ortho2} \end{equation}
Also, in analogy to eq.(\ref{vv}),  we have:
\begin{equation}
\left(  x_{i}^\mu - x_r^\mu \right) \left(  x_{j}^\nu - x_r^\nu \right) V^{(h_{ij})}_{ \mu\nu}(\alpha) = d(d-1)~ V(\alpha)~~~~~~~\forall~ r\neq i,j.       \label{vvv} \end{equation}

Notice also that $V^{(h_{ij})}_{ \mu\nu}(\alpha)$ 
is entirely contained in the two-dimensional subspace spanned by $n^{(\alpha\cap\alpha_i)}_{ \mu}(\alpha)$ and $n^{(\alpha\cap\alpha_j)}_{\mu}(\alpha)$. This base can be made orthonormal by defining:
\begin{equation}
 n^{(i)}_\mu(\alpha) = n^{(\alpha\cap\alpha_i)}_{ \mu}(\alpha);~~~~~n^{(j)}_\mu(\alpha) = \frac{1}{\sqrt{1-c^2}} n^{(\alpha\cap\alpha_j)}_{\mu}(\alpha) -  \frac{c}{\sqrt{1-c^2}} n^{(\alpha\cap\alpha_i)}_{\mu}(\alpha) \label{ortobase} \end{equation}
 where  $c=n^{(\alpha\cap\alpha_i)}_{ \mu}(\alpha)  g^{\mu\nu}(\alpha)   n^{(\alpha\cap\alpha_j)}_{\nu}(\alpha)$.
 The vectors $n^{(j)}_\mu(\alpha)$ and $n^{(i)}_\mu(\alpha)$ satisfy now orthonormality relation with respect to the metric $g_{\mu\nu}(\alpha)$:
\begin{equation}
n_{\mu}^{(a)}(\alpha) g^{\mu\nu}(\alpha) n_{\nu}^{(b)}(\alpha) = \delta^{ab}~~~~~~~~~~~~~a,b\in\{i,j\}. \label{onbase} \end{equation}
The covariant tensor of eq.(\ref{hinge}) can then be written as:
\begin{equation}
V^{(h_{ij})}_{ \mu_1\mu_2}(\alpha) = n_{ \mu_1\mu_2}^{(ij)}(\alpha) V(h_{ij})    \label{hinge2} \end{equation}
where
\begin{equation}
n_{ \mu_1\mu_2}^{(ij)}(\alpha) = \left( n_{\mu_1}^{(i)}(\alpha) n_{\mu_2}^{(j)}(\alpha) -n_{\mu_2}^{(i)}(\alpha) n_{\mu_1}^{(j)}(\alpha) \right) \label{nmunu} \end{equation}
and $V(h_{ij})$ is the absolute value\footnote{This implies that $V(h_{ij})$  is independent of the orientation: $V(h_{ij})=V(h_{ji}) > 0$} of the  $d-2$ dimensional volume of the hinge.

\section{Derivatives on a simplicial lattice.}
\label{derivativessl}
Derivatives are replaced on a lattice by finite differences. This is rather straightforward on regular hypercubic lattices which can be regarded as a discretization of a euclidean coordinate system where all coordinates are integer multiples of the lattice spacing. Regge Calculus on the other hand is defined on a simplicial complex which is in general not regular, and the  $d+1$ faces of each simplex point into different directions which are not related to any coordinate system. 

Defining on a simplicial lattice  the analogue  of the partial derivative $\partial_\mu$ with the further requirement that it transforms as  a covariant vector under the coordinate transformations defined in the previous sections is not a trivial problem and is the object of the present section.
To start with, we shall consider only derivatives of scalar quantities; derivatives of vectors and tensors need to be made covariant and require the notion of parallel transport. They will be discussed in the following sections.

Let  $\varphi_c(x)$ be a scalar field of the continuum theory  and $\varphi(\alpha)$  the corresponding field  on a simplicial lattice.  The partial derivative $\partial_\mu \varphi_c(x)$ transforms as a covariant vector, so we want to construct on the simplicial lattice  a new field $\hat{\partial}_\mu \varphi(\alpha)$ that transforms as a covariant vector under the coordinate transformations defined in (\ref{gctalpha}), depends on the value of $\varphi(\alpha)$ in the simplex $\alpha$ and in its neighbouring simplices and reduces to  $\partial_\mu\varphi_c(x)$ in the continuum limit. 

Let us denote by $\alpha_i$ with $i=1,2,\dots,d+1$ the $d+1$ simplices that have one face in common with $\alpha$. We shall use the following conventions: if $P_1,P_2,\dots,P_{d+1}$ are the vertices of $\alpha$ with coordinates $x_1^\mu,\dots  x_{d+1}^\mu$, then the simplex $\alpha_i$ denotes the simplex that has in common with $\alpha$ the $d-1$-dimensional face that does not contain the vertex $P_i$.

We then define the derivative of a scalar field $\varphi(\alpha)$ on a simplicial lattice as follows\footnote{We shall denote with $\hat{\partial}_\mu$ the partial derivative on the simplicial lattice to distinguish it from the one in the continuum $\partial_\mu$.}:
\begin{equation}
\hat{\partial}_\mu \varphi(\alpha) = \frac{1}{2} \sum_{i=1}^{d+1} \left[\varphi(\alpha_i) - \varphi(\alpha) \right] \frac{V_\mu^{(\alpha\cap\alpha_i)}(\alpha)}{V(\alpha)} \label{derivative} \end{equation}
where $V_\mu^{(\alpha\cap\alpha_i)}$, defined in (\ref{area}), is a covariant vector whose modulus is the $d-1$-dimensional volume $V(\alpha\cap\alpha_i)$ of the face $\alpha\cap\alpha_i$ and whose direction is orthogonal to $\alpha\cap\alpha_i$ (see eq.(\ref{nmu})). 

We now associate to the simplices  $\alpha$ and $\alpha_i$  a length $ l(\alpha|\alpha_i)$  defined as\footnote{Notice that $ l(\alpha|\alpha_i)$ is not symmetric: in general $ l(\alpha|\alpha_i) \neq  l(\alpha_i|\alpha) $.}
\begin{equation}
 l(\alpha|\alpha_i) = \frac{V(\alpha)}{V(\alpha\cap\alpha_i)} =\frac{n_\mu^{(\alpha\cap\alpha_i)}(\alpha)}{d} (x_{i}^\mu - x_j^\mu)~~~~~~~j\neq i,   \label{latticespacing} \end{equation} 
 then the derivative takes the natural form:
 \begin{equation}
 \hat{\partial}_\mu \varphi(\alpha) = \frac{1}{2} \sum_{i=1}^{d+1} \frac{ \left[\varphi(\alpha_i) - \varphi(\alpha) \right]}{ l(\alpha|\alpha_i)} n_\mu^{(\alpha\cap\alpha_i)}(\alpha) \label{derivative2} \end{equation}
where the length $ l(\alpha|\alpha_i)$ plays locally the role of a lattice spacing.

 Eq. (\ref{derivative}) and (\ref{derivative2}) can be further simplified by noticing that the area vectors $V_\mu^{(\alpha\cap\alpha_i)}$ of a simplex $\alpha$ are not linearly independent and satisfy the well known relation:
 \begin{equation}
 \sum_{i=1}^{d+1} V_\mu^{(\alpha\cap\alpha_i)} = 0. \label{sumarea0} \end{equation}
 By using (\ref{sumarea0}) we have then:
 \begin{equation}
 \hat{\partial}_\mu \varphi(\alpha) = \frac{1}{2} \sum_{i=1}^{d+1} \varphi(\alpha_i) \frac{V_\mu^{(\alpha\cap\alpha_i)}(\alpha)}{V(\alpha)}= \frac{1}{2} \sum_{i=1}^{d+1} \frac{\varphi(\alpha_i)}{ l(\alpha|\alpha_i)} n_\mu^{(\alpha\cap\alpha_i)}(\alpha). \label{derivative3}
 \end{equation}

Consider now a field $\varphi_c(x)$ of the continuum theory and define the field $\varphi(\alpha)$ on the simplicial complex as:
\begin{equation}
\varphi(\alpha)  = \varphi_c(\hat{x}(\alpha)) \label{simplexcenter} \end{equation}
where $\hat{x}^\mu(\alpha)$ are the coordinates of a point inside $\alpha$ that may be considered the ``center'' of $\alpha$\footnote{A possible choice is the circumcenter of $\alpha$, but this choice is not unique unless $\alpha$ is a regular simplex, which is the case considered below.}.

By inserting (\ref{simplexcenter}) into (\ref{derivative3}) and expanding around $\hat{x}^\mu(\alpha)$ we have:
 \begin{equation}
  \hat{\partial}_\mu \varphi(\alpha)=  \frac{1}{2} \sum_{i=1}^{d+1} \left( \hat{x}^\nu(\alpha_i) - \hat{x}^\nu(\alpha) \right) \partial_\nu  \varphi_c(\hat{x}(\alpha)) \frac{V_\mu^{(\alpha\cap\alpha_i)}(\alpha)}{V(\alpha)} + O\left( (\hat{x}^\nu(\alpha_i) - \hat{x}^\nu(\alpha))^2 \right). \label{contlimit} \end{equation}
 
 If the simplices $\alpha$ and $\alpha_i$ are generic, the r.h.s. of (\ref{contlimit}) cannot be calculated due to the ambiguity implicit in the choice of $\hat{x}^\nu(\alpha_i)$  and in general, even neglecting terms of second order in $ \hat{x}^\nu(\alpha_i) - \hat{x}^\nu(\alpha)$, the partial derivatives $ \hat{\partial}_\mu \varphi(\alpha)$ and $ \partial_\nu  \varphi_c(\hat{x}(\alpha)) $ will not coincide. However if all the simplices involved are regular, then the sum at the r.h.s. of (\ref{contlimit}) can be calculated\footnote{The explicit calculation is rather lengthly and is better done by choosing the same euclidean coordinates and metric in all the simplices involved.}  and gives:
 \begin{equation}
  \sum_{i=1}^{d+1} \left( \hat{x}^\nu(\alpha_i) - \hat{x}^\nu(\alpha) \right) V_\mu^{(\alpha\cap\alpha_i)}(\alpha) = 2 {V(\alpha)} \delta_\mu^\nu,  \label{sumi} \end{equation}
 which implies:
 \begin{equation}
  \hat{\partial}_\mu \varphi(\alpha) =  \partial_\mu  \varphi_c(\hat{x}(\alpha)) + O\left( |(\hat{x}(\alpha_i) - \hat{x}(\alpha))|^2 \right). \label{contlimit2} \end{equation}
  
It should also be noticed that in the case of regular hypercubic lattice, where $\alpha$ and $\alpha_i$ are $d$-dimensional hypercubes, $ l(\alpha|\alpha_i)$ coincides with the lattice spacing and the derivative (\ref{derivative3}) reduces to the usual symmetric finite difference operation on the lattice. So, in a sense, the derivative $\hat{\partial}_\mu \varphi(\alpha)$ is a generalization to a simplicial lattice of the symmetric finite difference on a cubic lattice .
 
 Given the form (\ref{derivative3}) for the derivative on a simplicial lattice it is immediate to write the action for a scalar field coupled with the metric tensor. In the continuum the action is:
 \begin{equation}
 S_{\varphi_c} = \int d^dx \sqrt{g(x)} \left[ g^{\mu\nu}(x) \partial_\mu \varphi_c(x) \partial_\nu \varphi_c(x) +{ \cal V}\left(\varphi_c (x)\right) \right]. \label{contscalar} \end{equation}
 where ${\cal V}$ is an arbitrary potential. 
 The corresponding simplicial action is simply obtained by replacing $\int d^dx \sqrt{g(x)}$ with  $\sum_{\alpha} V(\alpha)$, the continuum variable $x$ with the  label $\alpha$ of the simplex and the partial derivative $\partial_\mu$ with $\hat{\partial}_\mu$:
 \begin{equation}
 S_{\varphi} = \sum_{\alpha} V(\alpha)  \left[ g^{\mu\nu}(\alpha) \hat{\partial}_\mu \varphi(\alpha) \hat{\partial}_\nu \varphi(\alpha) +{ \cal V}\left(\varphi (\alpha)\right) \right].
 \label{simplscalar}
 \end{equation}
 The derivative $\hat{\partial}_\mu \varphi(\alpha)$ in (\ref{simplscalar}) can be now replaced by the r.h.s. of (\ref{derivative3}), and the kinetic term becomes:
 \begin{equation}
 S_{\varphi,kin} = \frac{1}{4} \sum_{\alpha}\sum_{i,j=1}^{d+1} \frac{ g^{\mu\nu}(\alpha) V_\mu^{(\alpha\cap\alpha_i)}(\alpha) V_\nu^{(\alpha\cap\alpha_j)}(\alpha)}{V(\alpha)} \varphi(\alpha_i) \varphi(\alpha_j).  \label{kinetics} \end{equation}
 
 The kinetic term (\ref{kinetics}) has a coupling between a simplex $\alpha_i$ and a simplex $\alpha_j$ that for $i\neq j$ have in common only a $d-2$ dimensional hinge (not a $d-1$ dimensional face). This is different from the actions for scalar fields on a simplicial lattice previously used in the literature, where either the scalar fields were defined on the sites of the simplectic lattice \cite{Hamber:1993gn}  \cite{McDonald:2010bj} or they were defined as in the present case on the simplices (the sites of the dual lattice)  but with couplings only between simplices with a face in common \cite{Li:2017uao}.
 
\section{Parallel Transport and Christoffel Symbol.}
\label{PT}

In order to proceed in analogy with the Einstein theory of gravity we have now to introduce the notion of parallel transport. Consider a contravariant vector $A^\mu(\alpha)$. According to (\ref{derivative}) the derivative of $A^\mu(\alpha)$ involves the differences  $A^\mu(\alpha)-A^\mu(\alpha_i)$, which however are not vectors since the two terms of the difference transform with different matrices, respectively  $\Lambda(\alpha)$ and $\Lambda(\alpha_i)$, under general coordinate transformations.

In order to define  covariant differences (and then a covariant derivative) that transforms like  vectors we need to introduce, as in the continuum case, the notion of parallel transport of a contravariant vector $ A^\mu(\beta)$ from a simplex $\beta$ onto a neighbouring simplex $\alpha$.  We define the transported vector $A_{(\alpha)}^\mu(\beta)$ as:
\begin{equation}
A^\mu(\beta)~~ \Longrightarrow ~~A_{(\alpha)}^\mu(\beta) = K^\mu_{\ \nu}(\alpha|\beta) A^\nu(\beta) \label{paralleltransp} \end{equation}
where $K^\mu_{\ \nu}(\alpha|\beta)$ is entirely determined by the following properties:
\begin{itemize}
\item If $g_{\mu\nu}(\alpha) =  g_{\mu\nu}(\beta)$ then $K^\mu_{\ \nu}(\alpha|\beta) = \delta^\mu_{\ \nu}$.
\item
$A_{(\alpha)}^\mu( \beta)$ transforms  as a contravariant vector in $\alpha$, namely it transforms with  $ \Lambda^\mu_{\ \nu}(\alpha)$ under general coordinate transformations:
\begin{equation}
A_{(\alpha)}^{'\mu}(\beta) = \Lambda^\mu_{\ \nu}(\alpha) A_{(\alpha)}^\nu(\beta). \label{ptr}
\end{equation}
\end{itemize}
The matrix  $ K^\mu_{\ \nu}(\alpha|\beta)$ does not transform as a tensor but rather as link variable on the dual lattice. In fact from (\ref{ptr}) and (\ref{paralleltransp}) one easily finds:
\begin{equation}
K^{'\mu}_{\ \ \nu}(\alpha|\beta) = \Lambda^\mu_{\ \rho}(\alpha) K^{\rho}_{\ \sigma}(\alpha|\beta)\Lambda^{-1 \sigma}_{\ \ \ \ \nu}(\beta).  \label{link} \end{equation}
Notice that if we start in (\ref{link}) from a coordinate system where $g_{\mu\nu}(\alpha)=g_{\mu\nu}(\beta)$, then $ K^{\rho}_{\ \sigma}(\alpha|\beta) = \delta^\rho_{\ \sigma}$ and in a generic coordinate system $K^{'\mu}_{\ \ \nu}(\alpha|\beta)$ can always be written in the form:
\begin{equation}
K^{'\mu}_{\ \ \nu}(\alpha|\beta) = \Lambda^\mu_{\ \rho}(\alpha) \Lambda^{-1 \rho}_{\ \ \ \ \nu}(\beta).  \label{link2} \end{equation}
It follows from (\ref{link2}) that $K^{'\mu}_{\ \ \nu}(\beta|\alpha) $ is the inverse of $K^{'\mu}_{\ \ \nu}(\alpha|\beta) $:
\begin{equation}
K^{'\mu}_{\ \ \nu}(\alpha|\beta) K^{'\nu}_{\ \ \rho}(\beta|\alpha)  = \delta^\mu_{\ \rho}. \label{inverse} \end{equation}
Scalar quantities are obviously invariant under parallel transform. This determines, together with (\ref{paralleltransp}), the parallel transport of a covariant vector:
\begin{equation}
A_\mu(\beta)~~ \Longrightarrow ~~A_{(\alpha)\mu}(\beta) = A_\nu(\beta) K_{\ \mu}^\nu(\beta|\alpha).  \label{covtr} \end{equation}
Eq.s (\ref{paralleltransp}) and (\ref{covtr}) can easily be generalised to tensors of arbitrary rank: in particular it follows from the definition of parallel transport that the parallel transport of $g_{\mu\nu}(\beta)$ to a neighbour simplex $\alpha$ coincides with $g_{\mu\nu}(\alpha)$, so that the following identity holds:
\begin{equation}
g_{\mu\nu}(\alpha) =  g_{\rho\sigma}(\beta)  K_{\ \mu}^\rho(\beta|\alpha) K_{\ \nu}^\sigma(\beta|\alpha).\label{gtransp} \end{equation}
The last equation defines $K_{\ \mu}^{\rho}(\alpha|\beta) $ implicitely as a function of the metric tensor, but unlike the continuum case it is quadratic in  $K_{\ \mu}^{\rho}(\alpha|\beta) $. Therefore  $K_{\ \mu}^{\rho}(\alpha|\beta) $ cannot be expressed, as in the continuum, by linear combinations of the derivatives of the metric tensor unless the equation  is linarized by neglecting higher order terms in the lattice constant $l(\alpha|\beta)$ (see discussion below).

In a hypothetical  first order formulation, analogue of the Palatini formalism of the continuum theory,  $ K_{\ \mu}^\rho(\alpha|\beta)$ and $g_{\mu\nu}(\alpha)$ would be  treated as independent dynamical variables and eq. (\ref{gtransp}) should arise from the eq.s of motion.  
We are not going to discuss this formulation in the present paper.

The variation of $ A^\mu(\beta)$ as a result of the parallel transport from $\beta$ to $\alpha$ is then given by:
\begin{equation}
\delta_{\beta\rightarrow\alpha}A^\mu(\beta) = A_{(\alpha)}^\mu(\beta) - A^\mu(\beta) = \left[ K^\mu_{\ \nu}(\alpha|\beta)-\delta^\mu_{\ \nu} \right] A^\nu(\beta). \label{ptvar} \end{equation}
Notice that  in (\ref{ptvar}) only the component of $A^\nu(\beta)$ orthogonal to the face $\alpha\cap\beta$ contributes to the variation. In fact from (\ref{link2}) and (\ref{commonLa}) we have:
\begin{equation}
\left[ K^\mu_{\ \nu}(\alpha|\beta)-\delta^\mu_{\ \nu} \right] \left(x_i^\nu - x_j^\nu \right)=0~~~~~~~~i,j\in \alpha\cap\beta.   \label{parcr} \end{equation}

Consider now a contravariant vector $A^\mu(\alpha)$. The covariant difference between two neighbouring simplices $\alpha$ and $\beta$ is defined as:
\begin{equation}
\hat{D}A^\mu(\alpha|\beta) = A_{(\alpha)}^\mu(\beta) -  A^\mu(\alpha)  =  A^\mu(\beta) - A^\mu(\alpha) + \delta_{\beta\rightarrow\alpha}A^\mu(\beta) \label{covdiff} \end{equation}
and it transforms as a contravariant vector in $\alpha$:
\begin{equation}
\hat{D}A^\mu(\alpha|\beta)' = \Lambda^\mu_\rho(\alpha)~ \hat{D}A^\rho(\alpha|\beta). \label{covtrans} \end{equation}
The covariant difference is not antisymmetric under exchange of $\alpha$ and $\beta$, but rather it satisfies the relation:
\begin{equation}
\hat{D}A^\mu(\beta|\alpha) = -  K^\mu_{\ \nu}(\beta|\alpha)~ \hat{D}A^\nu(\alpha|\beta). \label{antisD} \end{equation}

The covariant derivative of a contravariant vector is obtained by replacing in eq. (\ref{derivative2}) the differences with the corresponding covariant differences :
\begin{equation}
\hat{D}_\mu A^\nu(\alpha) = \frac{1}{2} \sum_{i=1}^{d+1} \frac{K^\nu_\rho(\alpha|\alpha_i) A^\rho(\alpha_i) - A^\nu(\alpha)}{l(\alpha|\alpha_i)} n_\mu^{(\alpha\cap\alpha_i)}(\alpha) \label{covder} \end{equation}
where $l(\alpha|\alpha_i)$ is given by (\ref{latticespacing}).
As in the case of ordinary derivatives eq.(\ref{covder}) can be further simplified according to eq. (\ref{sumarea0}):
\begin{equation}
\hat{D}_\mu A^\nu(\alpha) = \frac{1}{2} \sum_{i=1}^{d+1} \frac{K^\nu_\rho(\alpha|\alpha_i) A^\rho(\alpha_i)}{l(\alpha|\alpha_i)}n_\mu^{(\alpha\cap\alpha_i)}(\alpha). \label{covder2} \end{equation}

The covariant derivative (\ref{covder2}) can be written as the sum an ordinary derivative plus a term involving a discrete analogue $\Gamma^{\nu}_{\mu\rho}(\alpha|\alpha_i)$ of the Christoffel symbol:
\begin{equation}
\hat{D}_\mu A^\nu(\alpha)  = {\hat\partial}_\mu A^\nu(\alpha) +  \frac{1}{2} \sum_{i=1}^{d+1} \Gamma^{\nu}_{\mu\rho}(\alpha|\alpha_i) A^\rho(\alpha_i), \label{covder3}
\end{equation}

where
\begin{equation}
 \Gamma^{\mu}_{\ \nu\rho}(\alpha|\alpha_i) = \frac{ K^\mu_{\ \rho}(\alpha|\alpha_i)-\delta^\mu_{\ \rho} }{l(\alpha|\alpha_i)} n^{(\alpha\cap\alpha_i)}_{ \nu}(\alpha). \label{Christ} \end{equation}
 Notice that according to eq.s (\ref{parcr})  $\Gamma^{\mu}_{\ \nu\rho}(\alpha|\alpha_i) $ is different from zero only when the index $\rho$ is orthogonal to the face $\alpha\cap\alpha_i$, hence the Christoffel symbol can be written in the form:
 \begin{equation}
 \Gamma^{\mu}_{\ \nu\rho}(\alpha|\alpha_i)  = \Gamma^{\mu}(\alpha|\alpha_i)  n^{(\alpha\cap\alpha_i)}_{ \nu}(\alpha)  n^{(\alpha\cap\alpha_i)}_{ \rho}(\alpha)
  \label{Christ2}
  \end{equation}
 where
 \begin{equation}
  \Gamma^{\mu}(\alpha|\alpha_i) =  \frac{ K^\mu_{\ \rho}(\alpha|\alpha_i)-\delta^\mu_{\ \rho} }{l(\alpha|\alpha_i)}  n^{\rho (\alpha\cap\alpha_i)}(\alpha). \label{Christ3} \end{equation}
  The symmetry of $ \Gamma^{\mu}_{\ \nu\rho}(\alpha|\alpha_i) $ in the indices $\nu$ and $\rho$ is obvious from eq.(\ref{Christ2}).

In the continuum theory  the Christoffel symbol can be expressed in terms of the derivatives of the metric tensor. We can try to do the same thing here by writing  eq.(\ref{gtransp}) in terms of the Christoffel symbol (\ref{Christ}).  We find:
\begin{eqnarray}
\frac{g_{\mu\nu}(\alpha) - g_{\mu\nu}(\beta)}{l(\alpha|\beta)}& = n^{(\alpha\cap\beta)\tau}(\alpha) \left[\Gamma_{\ \mu\tau}^{\rho}(\alpha|\beta) g_{\rho\nu}(\beta) + \Gamma_{\ \nu\tau}^{\rho}(\alpha|\beta) g_{\rho\mu}(\beta)\right]  \nonumber \\& +l(\alpha|\beta)  n^{(\alpha\cap\beta)\tau}(\alpha) n^{(\alpha\cap\beta)\tau'}(\alpha) \Gamma_{\ \mu\tau}^{\rho}(\alpha|\beta)\Gamma_{\ \nu\tau'}^{\sigma}(\alpha|\beta) g_{\rho\sigma}(\beta).
\label{qq} \end{eqnarray}
The l.h.s of (\ref{qq})  is essentially  the derivative of the metric tensor along a direction orthogonal to the face $\alpha\cap\beta$. The r.h.s consists of a linear term, that resembles the one of the continuum theory, and of a quadratic term which however is  of order $l(\alpha|\beta)$  and hence vanishes in the continuum limit.

On the lattice the Christoffel symbol depends only on the $d$ components of $ \Gamma^{\mu}(\alpha|\beta)$ as defined in (\ref{Christ3}), and it is then convenient to express (\ref{qq}) in terms of $ \Gamma^{\mu}(\alpha|\beta)$:
\begin{eqnarray}
\frac{g_{\mu\nu}(\alpha) - g_{\mu\nu}(\beta)}{l(\alpha|\beta)}& = n^{(\alpha\cap\beta)}_\mu(\alpha)g_{\rho\nu}(\beta)  \Gamma^{\rho}(\alpha|\beta) + n^{(\alpha\cap\beta)}_\nu(\alpha)g_{\rho\mu}(\beta)  \Gamma^{\rho}(\alpha|\beta)  \nonumber \\& +l(\alpha|\beta)  n^{(\alpha\cap\beta)}_\mu(\alpha) n^{(\alpha\cap\beta)}_\nu(\alpha)  g_{\rho\sigma}(\beta)  \Gamma^{\rho}(\alpha|\beta) \Gamma^{\sigma}(\alpha|\beta). 
\label{qq2} \end{eqnarray}
If we contract eq.(\ref{qq2}) with $ n^{(\alpha\cap\beta)\mu}(\alpha)$ and with $\left(x_i^\nu - x_j^\nu \right)$ ($i,j\in \alpha\cap\beta$), then eq.(\ref{qq2}) becomes linear:
\begin{equation}
  n^{(\alpha\cap\beta)\mu}(\alpha)\left(x_i^\nu - x_j^\nu \right)\frac{g_{\mu\nu}(\alpha) - g_{\mu\nu}(\beta)}{l(\alpha|\beta)} = \left(x_i^\nu - x_j^\nu \right)g_{\nu\rho}(\beta)  \Gamma^{\rho}(\alpha|\beta). \label{qqq} \end{equation}
  
  This equation shows that all components of $g_{\nu\rho}(\beta)  \Gamma^{\rho}(\alpha|\beta) $ with the index $\nu$ belonging to the $d-1$ dimensional subspace $\alpha\cap\beta$ can be expressed also on the lattice as derivatives of the metric tensor.  
Instead the perpendicular component $ n^{(\alpha\cap\beta)\nu}(\alpha)g_{\nu\rho}(\beta)  \Gamma^{\rho}(\alpha|\beta) $ is solution of the quadratic equation
\begin{eqnarray} n^{(\alpha\cap\beta)\mu}(\alpha) n^{(\alpha\cap\beta)\nu}(\alpha)\frac{g_{\mu\nu}(\alpha) - g_{\mu\nu}(\beta)}{l(\alpha|\beta)}& = 2 n^{(\alpha\cap\beta)\nu}(\alpha) g_{\nu\rho}(\beta)  \Gamma^{\rho}(\alpha|\beta) \nonumber \\& +l(\alpha|\beta)  g_{\rho\sigma}(\beta)  \Gamma^{\rho}(\alpha|\beta) \Gamma^{\sigma}(\alpha|\beta) \label{qIV} \end{eqnarray}
which becomes linear only in the limit $l(\alpha|\beta) \rightarrow 0$.

We conclude this section with the proof that the divergence theorem for an arbitrary contravariant vector $A^\mu(\alpha)$, defined on a $d$-dimensional simplicial complex with boundary, is exactly satisfied. 

In the continuum the divergence theorem states that given a contravariant vector $A^\mu(x)$ on a $d$-dimensional manifold ${\mathcal M}$ the following identities hold:
\begin{equation}
\int_{\mathcal M} dx \sqrt{g(x)} D_\mu A^\mu(x)  = \int_{\partial {\mathcal M}}dx \sqrt{h(x)} n_\mu(x) A^\mu(x) \label{divergence} \end{equation}
where $h(x)$ is the determinant of the metric on $\partial {\mathcal M}$ induced by pulling back the metric from $ {\mathcal M}$ and $n_\mu(x)$ is the unit vector orthogonal to $\partial {\mathcal M}$ in $x$.

We consider now a triangulation of ${\mathcal M}$, namely a simplicial complex $\hat{\mathcal M}$ whose boundary $\partial \hat{\mathcal M}$  is made of $d-1$ dimensional simplices. Each simplex in  $\partial \hat{\mathcal M}$ is a  face of some simplex in $\hat{\mathcal M}$. 
We also associate to  the boundary $\partial \hat{\mathcal M}$ a layer of $d$-dimensional simplices, which we shall  denote $\hat{\mathcal M}_B$, defined as the ensamble of simplices of $\hat{\mathcal M}$ which have at least one face belonging to the boundary $\partial \hat{\mathcal M}$.

We can now write the simplicial analogue of the l.h.s of (\ref{divergence}) as:
\begin{equation} 
\sum_{\alpha\in \hat{{\mathcal M}}} V(\alpha) \hat{D}_\mu A^\mu(\alpha) = \frac{1}{2} \sum_{\alpha,\beta \in \hat{{\mathcal M}}}  \left[ K^\mu_{\ \rho}(\alpha|\beta) A^\rho(\beta)- A^\mu(\alpha)\right] V_\mu^{(\alpha\cap\beta)}(\alpha) \label{lattdiv} \end{equation}
where the sum at the r.h.s. is understood to extend over all pairs of simplices $\alpha$ and $\beta$ that have a $d-1$ dimensional face in common.
We can now use on the r.h.s. of (\ref{lattdiv}) the  identity
\begin{equation}
V_\mu^{(\alpha\cap\beta)}(\alpha) K^\mu_{\ \rho}(\alpha|\beta) = 
-V_\rho^{(\alpha\cap\beta)}(\beta) 
\label{Vtransport} \end{equation}
 which holds for any pair  of neighbouring simplices $\alpha$ and $\beta$. Eq. (\ref{Vtransport})  can be easily proved by just going from generic  coordinates to a choice of coordinates where $g_{\mu\nu}(\alpha)=g_{\mu\nu}(\beta)$.

By applying (\ref{Vtransport}) in (\ref{lattdiv}) one finds that the two terms at the r.h.s. of (\ref{lattdiv}) become identical, modulo an irrelevant exchange of the labels $\alpha$ and $\beta$. We have:
\begin{equation}
\sum_{\alpha\in \hat{{\mathcal M}}} V(\alpha) \hat{D}_\mu A^\mu(\alpha) = - \sum_{\alpha\in \hat{{\mathcal M}}} A^\mu(\alpha) \sum_{\beta\in\hat{{\mathcal M}}}  V_\mu^{(\alpha\cap\beta)}(\alpha)
\label{lattdiv2} \end{equation}
where the sum over $\beta$  only extends  to the simplices  in $\hat{\mathcal M}$ that have a face in common with $\alpha$. If none of the faces of $\alpha$ belongs to the boundary $\partial \hat{\mathcal M}$, namely if $\alpha\notin \hat{\mathcal M}_B$, then the sum over $\beta$ at the r.h.s. of (\ref{lattdiv2}) vanishes identically (see eq.(\ref{sumarea0})). 

On the other hand if $\alpha\in \hat{\mathcal M}_B$, namely if one (or more) of the faces of $\alpha$ belong to  $\partial\hat{\mathcal M}$, then the sum over $\beta$ at the r.h.s. of (\ref{lattdiv2}) does not vanish and is given by the sum\footnote{In general the sum is made of a single term, but it is possible for a simplex to have more than one boundary face.} (with the sign changed) of the terms missing in the sum over $\beta$, which obviously correspond to $d-1$ dimensional simplices belonging to the boundary $\partial\hat{\mathcal M}$.

In order to write the final form of eq.(\ref{lattdiv2}) it is convenient to introduce some new notation. Let us denote the $d-1$ dimensional simplices belonging to $\partial\hat{\mathcal M}$ with $\bar{\alpha}$, $\bar{\beta}$, etc. and the corresponding $d$ dimensional simplices by the same greek letters without bar: $\bar{\alpha}\subset \alpha$, $\bar{\beta}\subset \beta$, etc. 
We can also define the contravariant vector $A^\mu$ on the boundary by simply putting:
\begin{equation}
A^\mu(\bar{\alpha}) = A^\mu(\alpha) \ \ \ \ \ \ \  \bar{\alpha} \in \alpha.         \label{pullback} \end{equation}
 Then we can write eq.(\ref{lattdiv2}) as:
\begin{equation}
\sum_{\alpha\in \hat{{\mathcal M}}} V(\alpha) \hat{D}_\mu A^\mu(\alpha) = \sum_{\bar{\alpha}\in \partial\hat{\mathcal M}} A^\mu(\bar{\alpha}) V_\mu^{(\bar{\alpha})}(\alpha) = \sum_{\bar{\alpha}\in \partial\hat{\mathcal M}} V(\bar{\alpha})\  n_\mu^{(\bar{\alpha})}(\alpha) A^\mu(\bar{\alpha})  \label{lattdiv3} \end{equation}
where $V(\bar{\alpha})$ is the $d-1$ dimensional volume of $\bar{\alpha}$.
Eq.(\ref{lattdiv3}) is the divergence theorem on a simplicial lattice.
It is remarkable that it is an exact result on the lattice and that it  has a precise correspondence with the continuum case given in eq.(\ref{divergence}).

\section{Riemann Curvature Tensor, Bianchi Identities and Einstein Action}
\label{Riemannsect}
We shall now introduce the Riemann  curvature tensor, which in a piecewise flat simplicial manifold is localized on the $d-2$ dimensional hinges.

Following the notation introduced at the end of Sec.\ref{sectens}, given a simplex $\alpha$ and two neighbouring simplices $\alpha_i$ and $\alpha_j$ we denote by $h_{ij}$ the hinge that they have in common. 

We now define a closed path $\gamma_{h_{ij}}$ that starting from $\alpha$ goes all round the hinge $h_{ij}$, more precisely:
\begin {equation}
\gamma_{h_{ij}} \equiv  \alpha\rightarrow\alpha_i\rightarrow\beta_1\rightarrow\dots\rightarrow\beta_h\rightarrow\alpha_j\rightarrow\alpha    \label{closedpath} \end{equation}
where $\beta_r$ with $r=1,2,\dots h$ are the other simplices that have $h_{ij}$ as a hinge. 
Notice that $\gamma_{h_{ij}}$ corresponds to a plaquette in the dual lattice and that   $\gamma_{h_{ji}}$ denotes the same path taken in opposite direction.
 
Given a contravariant vector $A^\mu(\alpha)$ we can define, according to the definitions of the previous section, the parallel transport of   $A^\mu(\alpha)$  around the hinge $h_{ji}$ along  $\gamma_{h_{ij}}$  starting and arriving in $\alpha$.
The variation of $A^\mu(\alpha)$ under  parallel transport along  $\gamma_{h_{ij}}$ is then given by:
\begin{equation}
\delta_{\gamma_{h_{ij}}} A^\mu(\alpha) =R^\mu_{\ \rho}(\gamma_{h_{ij}})  A^\rho(\alpha)  \label{curv1} \end{equation}
with
\begin{equation}
R^\mu_{\ \rho}(\gamma_{h_{ij}}) =  K^\mu_{\ \nu_1}(\alpha|\alpha_i) K^{\nu_1}_{\ \nu_2}(\alpha_i|\beta_1) \dots K^{\nu_{h}}_{\ \nu_{h+1}}(\beta_h|\alpha_j) K^{\nu_{h+1}}_{\ \rho}(\alpha_j|\alpha)- \delta^\mu_{\ \rho}.  \label{curv2} \end{equation}
From the transformation property of $K^\mu_{\ \nu}(\alpha|\beta)$ given in (\ref{link}) it follows immediately that $R^\mu_{\ \rho}(\gamma_{h_{ij}}) $ transforms as a mixed tensor with one covariant and one contravariant index.
The index $\mu$ in $R^\mu_{\ \rho}(\gamma_{h_{ij}}) $ can be lowered to define a covariant tensor of rank two:
\begin{equation}
R_{\mu \rho}(\gamma_{h_{ij}}) = g_{\mu\nu}(\alpha) R^\nu_{\ \rho}(\gamma_{h_{ij}}). \label{rank2} \end{equation}
The curvature tensor $R_{\mu \rho}(\gamma_{h_{ij}}) $ satisfies the symmetry relation:
\begin{equation}
R_{\mu \rho}(\gamma_{h_{ij}}) = R_{\rho \mu }(\gamma_{h_{ji}}) \label{Rsym} 
\end{equation}
where $\gamma_{h_{ji}}$ is the same path as $\gamma_{h_{ij}}$ but taken in the opposite  direction.
This property follows from an analogue property of $ K^\mu_{\ \nu}(\alpha|\beta)$, namely:
 \begin{equation}
K_{\mu \nu}(\alpha|\beta) =K_{\nu \mu}(\beta|\alpha) 
\label{Ksym}
\end{equation}
where the metric tensor has been used to lower indices in $K$ and eq.s (\ref{gtransp}) and (\ref{inverse}) have been applied.
Repeated use of (\ref{Ksym}) leads to eq.(\ref{Rsym}).

In eq.(\ref{curv1}) only the components of $A^\mu(\alpha)$ orthogonal to the hinge are modified under  parallel transport along  $\gamma_{h_{ij}}$, that is the only non vanishing elements of the curvature matrix $R^\mu_{\ \rho}(h_{ij})$ are the ones where the  index $\rho$ is orthogonal to the hinge $h_{ij}$. This is a direct consequence of eq.(\ref{parcr}) and thanks to the symmetry (\ref{Rsym}) this property applies to both covariant indices in $R_{\mu \rho}(\gamma_{\gamma_{h_{ij}}})$:
\begin{equation}
R_{\mu \rho}(\gamma_{h_{ij}}) (x_r^\rho - x_s^\rho) = (x_r^\mu - x_s^\mu)R_{\mu \rho}(\gamma_{h_{ij}}) = 0~~~~~~~~r,s\in h_{ij} \rightarrow r,s \neq i,j.
\label{curv3} \end{equation}
The curvature $R_{\mu \rho}(\gamma_{h_{ij}})$ is then entirely contained in the two dimensional subspace spanned by the orthonormal base vectors $n_{\mu}^{(i)}(\alpha)$ and $n_{\mu}^{(j)}(\alpha)$ introduced in (\ref{ortobase}). Since the orthonormality relations are preserved under parallel transport, the effect of a parallel transport along $\gamma_{h_{ij}}$ can only be a rotation by an angle $\theta_{ij}$ of the orthonormal base vectors in this two dimensional subspace. The angle $\theta_{ij}$ can  be identified as the deficit angle of the Regge calculus associated to the hinge $h_{ij}$.

An explicit expression for $R_{\mu \rho}(\gamma_{h_{ij}})$ in terms of the deficit angle and of the vectors $n_{\mu}^{(a)}(\alpha)$ ($a\in\{i,j\}$) can then be written, and reads:
\begin{equation}
R_{\mu\rho}(\gamma_{h_{ij}}) = \left( \cos \theta_{ij} - 1 \right) \sum_{a\in \{ij\}} n_{\mu}^{(a)}(\alpha) n_{\rho}^{(a)}(\alpha) + \sin \theta_{ij} ~ n_{\mu\rho}^{(ij)}(\alpha)  \label{curv4} \end{equation}
where $ n_{\mu\rho}^{(ij)}(\alpha) $ is given in (\ref{nmunu}).
Notice that the first term at the r.h.s. of (\ref{curv4}), which is of order $\theta_{ij}^2$ for small deficit angles, is independent of the orientation of the hinge, whereas the second term (order $\theta_{ij}$) changes sign if the orientation of the hinge is reversed. However eq. (\ref{curv4}) is consistent with (\ref{Rsym}) because $ n_{\mu\rho}^{(ij)}(\alpha)$ is antisymmetric in both pairs of indices $ij$ and $\mu\rho$.

The term in $\sin \theta_{ij}$ is the relevant one in Regge Calculus and for that reason we shall use the antisymmetric combination
\begin{equation}
R^{(-)}_{\mu\rho}(\gamma_{h_{ij}}) = \frac{1}{2} \left[  R_{\mu\rho}(\gamma_{h_{ij}}) -R_{\mu\rho}(\gamma_{h_{ji}}) \right] =  \sin \theta_{ij} ~ n_{\mu\rho}^{(ij)}(\alpha).  \label{antR} \end{equation}
As shown below the use of  $R^{(-)}_{\mu\rho}(\gamma_{h_{ij}})$ in place of $R_{\mu\rho}(\gamma_{h_{ij}})$, besides eliminating the higher order term in $\cos \theta_{ij}$, leads to a Riemann tensor which is independent of orientation of the hinge, an important feature for an unambiguous definition of the lattice action.

The first set of indices of the Riemann tensor can be identified with the two covariant indices in $R^{(-)}_{\mu\rho}(\gamma_{h_{ij}})$ and describe the rotation of a vector under parallel transport aroung the loop $\gamma_{h_{ij}}$.  The second set of indices describe the spacial orientation of the loop. They are contracted with the area element (in the continuum: $dx^\mu \wedge dx^\nu$) and on the simplicial lattice they should be orthogonal to the hinge $h_{ij}$ and hence proportional to $V^{(h_{ij})}_{ \mu\nu}(\alpha)$ as defined in (\ref{hinge}).  
In conclusion, the curvature tensor with four covariant indices should have the form:
\begin{equation}
{\cal R}_{\mu\nu,\rho\sigma}(\gamma_{h_{ij}}) = \frac{R^{(-)}_{\mu\nu}(\gamma_{h_{ij}})}{v(h_{ij})} V^{(h_{ij})}_{ \rho\sigma}(\alpha)  = \frac{\sin \theta_{ij} V(h_{ij})}{v(h_{ij})} n^{(ij)}_{ \mu\nu}(\alpha) n^{(ij)}_{ \rho\sigma}(\alpha).  \label{riemann} \end{equation}
For dimensional reasons the quantity $v(h_{ij})$ is a $d$-volume and can be identified with the support volume of the hinge $h_{ij}$. A precise and rigorous definition of the support volume can be found in \cite{Dec}, it will suffice here to know that a point $P$ of the simplicial complex belongs to the support of a hinge $h_{ij}$ if its minimal distance from a point of  $h_{ij}$ is less than that from any other hinge of the complex. 
It is also useful, as we shall see later on in this section, to define the volume $v(h_{ij}|\alpha)$, namely the volume of the part of the support of $h_{ij}$ that belongs to a given simplex $\alpha$. The following relations then obviously hold:
\begin{eqnarray}
\sum_{\alpha | \alpha \ni h_{ij}} v(h_{ij}|\alpha) &=& v(h_{ij}),   \label{alphav} \\
\sum_{h_{ij} | h_{ij} \in \alpha} v(h_{ij}|\alpha)&=& V(\alpha). \label{valpha} \end{eqnarray}

It is easy to check that the usual algebraic symmetries of the Riemann tensor are identically satisfied, namely the invariance under exchange of the two pairs of indices and the first Bianchi identity:
\begin{equation}
{\cal R}_{\mu\nu,\rho\sigma}(\gamma_{h_{ij}})+{\cal R}_{\mu\rho,\sigma\nu}(\gamma_{h_{ij}})+{\cal R}_{\mu\sigma,\nu\rho}(\gamma_{h_{ij}})=0.
\label{bianchi1} \end{equation}

We shall briefly discuss now, in the context of our approach, the second  Bianchi identity. This  is a differential identity, whose  formulation on a simplicial complex was already outlined in the original Regge paper \cite{Regge:1961px} and discussed in detail in \cite{Hamber:2001uq}.

First we prove that there is one Bianchi identity associated to each $d-3$ dimensional subsimplex, which we name $\sigma_{d-3}$, of the simplicial complex. Let us consider the dual lattice, namely the Voronoi tassellation generated by the vertices of the simplicial lattice. The dual of   $\sigma_{d-3}$ is a $3$-dimensional polytope (polyhedron) $\star\sigma_{d-3}$  whose faces are the plaquettes which are dual of the hinges that contain $\sigma_{d-3}$ as a subsimplex. 

Let $f$, $v$ and $s$ be respectively the number of faces, vertices and edges of the boundary of $\star\sigma_{d-3}$, which we shall assume has the topology of a sphere. Then the Euler relation holds, that we write as:
\begin{equation}
f = (s - v) + 2.     \label{Euler} \end{equation}
Each of the $v$ vertices correspond to a $d$-simplex, so there is an arbitrary coordinate choice attached to it. On the other hand a parallel transport matrix is associated to each of the $s$ edges (links), and by choosing the coordinate system in one of the simplices at the ends of the link the transport matrix can be made equal to the identity.
 This can be described as one link collapsing to a point with the two vertices at the ends becoming a single vertex. This new vertex does not correspond to a $d$-simplex anymore, but to the union of two simplices with the same metric.
This procedure can be repeated $v-1$ times, until there is only one vertex left, a coordinate transformation on this last vertex being just an overall transformation. 
If we denote by $s'$ the number of links left, namely the number of links where the parallel transport is non trivial, we have:
\begin{equation}
f = s' + 1.     \label{Euler2} \end{equation}
The curvatures (and the corresponding deficit angles) associated to each of the $f$ plaquettes (hinges) are obtained from products of transport matrices of the $s'$ links and since $f-s'=1$ they cannot be independent and must be related by one (and only one) identity (second Bianchi identity). 

In order to find a more explicit form for the second Bianchi identity let us follow the track of Regge's original paper. Let us denote  by $y_i ~ (i=1,\dots,f)$ the centers of the $f$ plaquettes in $\star\sigma_{d-3}$ \footnote{Each plaquette is dual to the hinge $h_i\equiv [x_1^\mu,\dots,x_{d-2}^\mu,y_i^\mu]$ where  $y_i^\mu$ is the additional vertex of $h_i$ that does not belong to $\sigma_{d-3}$ which is given in this notation by $\sigma_{d-3}\equiv [x_1^\mu,\dots,x_{d-2}^\mu]$. The center of the $i$-th plaquette corresponds to the position of $y_i^\mu$.}. By joining $y_i$ to $y_{i+1}$ modulo $f$ let us now construct a closed path $\gamma$ that divides the boundary of  $\star\sigma_{d-3}$ in two regions $\mathcal{M}$ and $\mathcal{M}'$. Let us choose in $\mathcal{M}$ (resp. $\mathcal{M}'$)  a vertex of $\star\sigma_{d-3}$ that corresponds to a given simplex $\alpha$ (resp. $\beta$) and denote by $\gamma_i$ a path that goes from $\alpha$ to $\beta$ along a sequence of links and crosses $\gamma$ in the section between $y_i$ and $y_{i+1}$. Let us now define $a_i = \gamma_i \gamma_{i+1}^{-1} $ and notice that $a_i$ is a closed path that starts and ends in $\alpha$ and encircles the $i$-th plaquette. 
Consider now the rotation matrix that describes the parallel transport along $a_i$:
\begin{equation}
S^\mu_{~\rho} (a_i) = K^\mu_{~\nu_1}(\alpha|\alpha^{(i)}_1)  K^{\nu_1}_{~\nu_2}(\alpha^{(i)}_1|\alpha^{(i)}_2)\dots  K^{\nu_{l_i}}_{~\rho}(\alpha^{(i)}_{l_i}|\alpha) = \delta^\mu_{~\rho} + R^\mu_{~\rho}(h_i|\alpha)  \label{loopai} \end{equation}
where $\alpha^{(i)}_k~(k=1,\dots,l_i)$ are the $k$-th simplices along the path $a_i$ and $R^\mu_{~\rho}(h_i|\alpha) $ is the curvature matrix associated  to the $i$-th plaquette. The latter is obtained by going around the $i$-th plaquette following the path $a_i$ starting end ending in $\alpha$; this is the same as going around the plaquette starting and ending in a simplex $\alpha'$ on the plaquette and then performing the parallel transport of the resulting curvature matrix from $\alpha'$ to $\alpha$ always following the path of $a_i$. 

The second Bianchi identity on the simplicial lattice is the given by the identity;
\begin{equation}
S^\mu_{~\nu_1}(a_1)S^{\nu_1}_{~\nu_2}(a_2)\dots S^{\nu_f}_{~\rho}(a_f) =\left( \delta^\mu_{~\nu_1} + R^\mu_{~\nu_1}(h_1|\alpha) \right) \dots \left( \delta^{\nu_f}_{~\rho} + R^{\nu_f}_{~\rho}(h_f|\alpha) \right) = \delta^\mu_\rho. \label{bid2}
\end{equation}
Notice that (\ref{bid2}) is not linear in the curvatures, and it becomes linear only in the limit of small curvatures, namely in the limit of small deficit angles, where it takes the form:
\begin{equation}
\sum_{i=1}^f  R^\mu_{~\rho}(h_i|\alpha) \approx 0.  \label{bid2lin} \end{equation}

Let us go back now to the curvature tensor ${\cal R}_{\mu\nu,\rho\sigma}(\gamma_{h_{ij}}) $ given in eq.(\ref{riemann}).
By contracting pairs of indices in ${\cal R}_{\mu\nu,\rho\sigma}(\gamma_{h_{ij}}) $ with the inverse metric $g^{\mu\rho}(\alpha)$ one obtains the Ricci tensor. This is given by:
\begin{equation}
{\cal R}_{\nu\sigma}(\gamma_{h_{ij}}) = \frac{\sin \theta_{h_{ij}} V(h_{ij}) \left( n_{\nu}^{(i)}(\alpha) n_{\sigma}^{(i)}(\alpha)  + n_{\nu}^{(j)}(\alpha) n_{\sigma}^{(j)}(\alpha) \right) }{2~ v(h_{ij})}.  \label{Ricci}  \end{equation}

By further contraction of the Ricci tensor with $g^{\nu\sigma}(\alpha)$ we obtain the curvature scalar:
\begin{equation}
{\cal R}(\gamma_{h_{ij}}) = \frac{\sin\theta_{h_{ij}} ~V(h_{ij})}{v(h_{ij})}. \label{curvsc} \end{equation}

The above expressions for the Riemann curvature tensor (\ref{riemann}), for the Ricci tensor (\ref{Ricci}) and for the curvature scalar (\ref{curvsc})  give the contribution to the curvature  coming  from a particular hinge $h_{ij}$.  

In order to have a direct correspondence with the continuum case we can define a Riemann tensor associated to each simplex $\alpha$ by taking a weighted sum over all the $\frac{d(d+1)}{2}$ hinges $h_{ij}$  that belong to $\alpha$:
\begin{equation}
{\cal R}_{\mu\nu,\rho\sigma}(\alpha) = \sum_{h_{ij}\in \alpha} \frac{v(h_{ij}|\alpha)}{V(\alpha)} {\cal R}_{\mu\nu,\rho\sigma}(\gamma_{h_{ij}})  \label{curvtens3} \end{equation}
where $v(h_{ij}|\alpha)$ has been defined in the discussion following eq.(\ref{riemann}) and the ratio $ \frac{v(h_{ij}|\alpha)}{V(\alpha)}$ corresponds to the fraction of the volume $V(\alpha)$ that belongs to the support of $h_{ij}$.

Similarly for the curvature scalar we can define:
\begin{equation}
{\cal R}(\alpha) =  \sum_{h_{ij} \in \alpha} \frac{v(h_{ij}|\alpha)}{V(\alpha)}  {\cal R}(\gamma_{h_{ij}}) = \frac{1}{V(\alpha)}   \sum_{h_{ij} \in \alpha}  \frac{v(h_{ij}|\alpha) V(h_{ij})}{v(h_{ij})}\sin\theta_{h_{ij}}. \label{curvsc2} \end{equation}

${\cal R}(\alpha)$ is the true lattice analogue of the curvature scalar $R(x)$ of Einstein continuum theory. Given the correspondence between the simplex volume $V(\alpha)$ (see eq.(\ref{simplexvolume})) and the integration volume of the continuum $\sqrt{g}~d^d x$ the lattice equivalent of the Einstein-Hilbert action is then:
\begin{equation}
S_l =  k \sum_\alpha   {\cal R}(\alpha)V(\alpha) =k \sum_h \sin\theta_h V(h) \label{latticeaction} \end{equation}
where $k$ is the Newton constant and the sum at the r.h.s. is over all hinges of the simplicial complex. Notice also that eq.(\ref{alphav}) has been used after exchanging summations in obtaining the last expression in (\ref{latticeaction}) which coincides with   Regge's  action in the limit of small deficit angles. The appearance of $\sin\theta_h$ in place of $\theta_h$ is not a new feature, and it seems a natural one when the analogy of gravity with gauge theories is made explicit on the lattice.  
\section{Higher Derivatives and Brans-Dicke Actions.} 
\label{HDBD}
Gravitational theories that  contain higher derivative terms, namely terms that are quadratic or of  higher order in the  curvature scalar or in the Riemann tensor, have received a lot of attention in recent years (see for instance \cite{Sotiriou:2008rp} and references therein).

In the original Regge Calculus the curvature is associated to the $d-2$ dimensional hinges of the simplicial complex. However we have shown in the previous section that the Riemann tensor can be associated to each simplex $\alpha$  by taking a suitable avarage over all the hinges belonging to $\alpha$ (see eq.(\ref{curvtens3})). 

This gives a straightforward prescription for writing on the simplicial lattice any arbitrary gravitational action in any dimension. It is sufficient to replace the $d$-dimensional invariant integration volume with the sum  of the  volume $V(\alpha)$ over all simplices and replace the Riemann tensor (and its contractions) with its discrete counterpart  (\ref{curvtens3}) on the simplicial lattice:
\begin{eqnarray} \int d^dx \sqrt{g(x)} &\Longrightarrow & \sum_\alpha V(\alpha), \label{cosmo}   \\  R_{\mu\nu\rho\sigma}(x) &\Longrightarrow & {\cal R}_{\mu\nu\rho\sigma}(\alpha). \label{correspondence} \end{eqnarray}

This correspondence has already  been shown in (\ref{latticeaction}) to reproduce Regge's original action\footnote{It is always understood that the deficit angle $\theta_h$ is replaced here by $\sin \theta_h$.}, and the r.h.s. of eq.(\ref{cosmo}) obviously gives  the simplicial version of the cosmological term.

In the case of gravitational actions with quadratic or higher order tems in the curvature the correspondence between the action in the  continuum  and the one on a simplicial lattice expressed in terms of the deficit angles $\theta_h$ is not unique and the prescription given in (\ref{correspondence}) provides a well defined and consistent way of constructing the lattice action.

 For instance for quadratic terms in the curvature eq.(\ref{correspondence})  gives:
\begin{eqnarray} \int   d^dx \sqrt{g(x)} R^2(x) &\Longrightarrow & \sum_\alpha V(\alpha) {\cal R}^2(\alpha)  \label{R2},   \\ \int   d^dx \sqrt{g(x)} R^{\mu\nu}(x)R_{\mu\nu}(x)  &\Longrightarrow &  \sum_\alpha V(\alpha) {\cal R}^{\mu\nu}(\alpha) {\cal R}_{\mu\nu}(\alpha), \label{Ricci2}\\ \int   d^dx \sqrt{g(x)} R^{\mu\nu\rho\sigma}(x)R_{\mu\nu\rho\sigma}(x)  &\Longrightarrow &  \sum_\alpha V(\alpha) {\cal R}^{\mu\nu\rho\sigma}(\alpha) {\cal R}_{\mu\nu\rho\sigma}(\alpha). \label{Riemann2}\end{eqnarray}

An explicit expression of the r.h.s. in (\ref{R2}), (\ref{Ricci2}) and (\ref{Riemann2}) in terms of the deficit angles can be easily obtained by replacing in them the Riemann curvature and its contractions as given in eq.(\ref{curvtens3}) and (\ref{curvsc2}). 

We are not interested here in the detailed expressions, except for remarking that they contain  mixed terms in the deficit angles of the form $\sin \theta_h \sin \theta_{h'}$ where $h$ and $h'$ are neighbouring hinges, namely hinges that have a simplex in common.

This is different from the simplest way of expressing  $R^2$ actions in terms of the deficit angles $\theta_h$, which would be to associate a factor $\sin^2 \theta_h$ (or $\theta^2_h$ in the small angle limit) to each hinge $h$ and sum, with suitable weights, over all hinges \cite{Hamber:1984kx}. Without mixed terms however all quadratic actions look the same when expressed in terms of the deficit angles. This difficulty was recognized already in \cite{Hamber:1985gx} where mixed term were introduced very much along the same lines as the ones presented here, namely by weighting the hinges in the curvature  proportionally to their support in the simplex.

$f(R)$ theories of  gravity, namely theories where the curvature scalar $R$ in Einstein action is replaced by an arbitrary function $f(R)$, can be written on the simplicial lattice by the usual replacement:
\begin{equation}
\int d^dx \sqrt{g(x)} f(R(x)) \Longrightarrow \sum_\alpha  V(\alpha)  f({\cal R}(\alpha)). \label{fR} \end{equation}
In the continuum  it is convenient to write the $f(R)$ gravity action as a linear action in $R$ by introducing an auxiliary scalar field $\Phi(x)$ (see the detailed discussion at page 12 in \cite{Sotiriou:2008rp}). The same applies, with the usual replacement rules, in the simplicial lattice case. The action at the r.h.s. of (\ref{fR}) is equivalent to
\begin{equation}
S_f =  \sum_\alpha  V(\alpha) \left[ \Phi(\alpha){ \cal R}(\alpha) - W(\Phi(\alpha)) \right] \label{fR2} \end{equation}
where the potential $W(\Phi)$ is related to the original function $f$ that appears in (\ref{fR}) by the equations:
\begin{eqnarray}{\cal R}(\alpha) &= & W'(\Phi(\alpha)) \label{W},   \\ f({\cal R}(\alpha))  &= &\Phi(\alpha) {\cal R}(\alpha) -  W(\Phi(\alpha)).  \label{W2}
\end{eqnarray}

The scalar field of eq.(\ref{fR2}) may be endowed with a kinetic term, following the  results of Section \ref{derivativessl}, and eq.(\ref{fR2}) with $W(\Phi(\alpha))=0$ becomes the simplicial lattice version of Brans-Dicke theory:
\begin{equation}
S_{BD} =  \sum_\alpha  V(\alpha)  \left[ \Phi(\alpha){ \cal R}(\alpha) - \frac{\omega}{\Phi(\alpha)} g^{\mu\nu}(\alpha) \hat{\partial}_\mu \Phi(\alpha)  \hat{\partial}_\nu \Phi(\alpha) \right].
\end{equation}

 \section{Coupling of gauge theories to gravity on a simplicial lattice.}
\label{sectiongauge}
Let us consider first a scalar field $\Phi^a(x)$ that transforms under a certain irreducible representation of a local symmetry group $G$. In the continuum a finite gauge transformation reads:
\begin{equation}
\Phi^a(x) \Rightarrow ( e^{i \eta^A(x) T_A} )^a_{~b} \Phi^b(x) \label{gaugetr} \end{equation}
where $\eta^A(x)$ are the local gauge parameters and $T_A$ the generators of the gauge group $G$.

On a simplicial lattice the space-time label $x$ is replaced by a  label $\alpha$ that runs over the simplices of the lattice. A gauge transformation can then be written by replacing $x$ with $\alpha$ everywhere in (\ref{gaugetr}):
\begin{equation}
\Phi^a(\alpha) \Rightarrow ( e^{i \eta^A(\alpha) T_A} )^a_{~b} \Phi^b(\alpha). \label{gaugetrsimpl} \end{equation}
Notice that the simplices $\alpha$ are the sites of the dual lattice, hence the gauge transformation (\ref{gaugetrsimpl}) is local in the dual lattice as it is usual in lattice gauge theories.

The lattice derivative of $\Phi^a(\alpha)$, as defined in (\ref{derivative}) or (\ref{derivative3}), does not transform according to (\ref{gaugetrsimpl}) and has to be replaced by a covariant derivative $\hat{D}_\mu$ defined as:
\begin{equation}
\hat{D}_\mu \Phi^a(\alpha) = \frac{1}{2} \sum_{i=1}^{d+1} U(\alpha|\alpha_i)^a_{~b} \Phi^b(\alpha_i) \frac{V_\mu^{(\alpha\cap\alpha_i)}(\alpha)}{V(\alpha)}  \label{gaugecovder} \end{equation}
where $U(\alpha|\alpha_i)$ is an element of the gauge group $G$, defined as usual on the link $(\alpha,\alpha_i)$ of the dual lattice, and transforms under a gauge transformation (\ref{gaugetrsimpl}) as:
\begin{equation}
U(\alpha|\alpha_i)^a_{~b} \Rightarrow ( e^{i \eta^A(\alpha) T_A} )^a_{~c}U(\alpha|\alpha_i)^c_{~f}  ( e^{-i \eta^A(\alpha_i) T_A} )^f_{~b}.  \label{Utransf} \end{equation}
The transformation properties of the covariant derivative (\ref{gaugecovder}) follow directly from (\ref{Utransf}):
\begin{equation}
\hat{D}_\mu \Phi^a(\alpha)  \Rightarrow ( e^{i \eta^A(\alpha) T_A} )^a_{~b} \hat{D}_\mu \Phi^b(\alpha).  \label{Dtransf} \end{equation}
The covariant derivative (\ref{gaugecovder}) can be written as the sum of an ordinary derivative and of a term containing the lattice equivalent of the gauge field $A_\mu(x)$:
\begin{equation}
\hat{D}_\mu \Phi^a(\alpha)  = \hat{\partial}_\mu \Phi^a(\alpha) + \sum_{i=1}^{d+1} {\cal A}^{~a}_{\mu~ b}(\alpha|\alpha_i)  \Phi^b(\alpha_i)
\label{gcovd} \end{equation}
where according to (\ref{gaugecovder}) the gauge field $ {\cal A}^{~a}_{\mu~ b}(\alpha|\alpha_i) $ is given by:
\begin{equation}
 {\cal A}^{~a}_{\mu~ b}(\alpha|\alpha_i)= \left[ U(\alpha|\alpha_i)^a_{~b} - \delta^a_{~b} \right] \frac{V_\mu^{(\alpha\cap\alpha_i)}(\alpha)}{V(\alpha)}.  \label{Amu} \end{equation}
Notice that the gauge field (\ref{Amu}) has a link nature and is not a function of the simplex $\alpha$, as one would expect from a naive correspondence with the continuum field $A_\mu(x)$, but of the $d-1$ dimensional face $\alpha\cap\alpha_i$ which in the dual lattice is the link joining the simplex $\alpha$ to $\alpha_i$.

Let $\bar{\Phi}_a(\alpha)$ be the conjugate scalar field of $ \Phi^a(\alpha)$. The gauge transformation of $\bar{\Phi}_a(\alpha)$ and its covariant derivative are obviously given by:
\begin{eqnarray} \bar{\Phi}_a(\alpha) &\Longrightarrow &  \bar{\Phi}_b(\alpha) ( e^{-i \eta^A(\alpha) T_A} )^b_{~a},\label{pbgt}   \\ \hat{D}_\mu \bar{\Phi}_a(\alpha)  & = &  \frac{1}{2} \sum_{i=1}^{d+1} \bar{\Phi}_b(\alpha_i)  U(\alpha_i|\alpha)^b_{~a}\frac{V_\mu^{(\alpha\cap\alpha_i)}(\alpha)}{V(\alpha)}.  \label{barcovder}\end{eqnarray}

We can now write the action for the kinetic term of $\Phi_a(\alpha)$ which is both invariant under gauge transformations and general coordinate transformations:
\begin{equation}
S_{kin} = \sum_\alpha V(\alpha) g^{\mu\nu}(\alpha)  \hat{D}_\mu \bar{\Phi}_a(\alpha)  \hat{D}_\nu \Phi^a(\alpha). \label{kinphi} \end{equation}
A more explicit expression for $S_{kin}$ can be obtained by inserting in (\ref{kinphi}) the explicit form of the covariant derivatives:
\begin{equation}
S_{kin} =\frac{1}{4} \sum_\alpha \sum_{i,j=1}^{d+1} V(\alpha) g^{ij}(\alpha) \bar{\Phi}_b(\alpha_i) \frac{ U(\alpha_i|\alpha)^b_{~c} U(\alpha|\alpha_j)^c_{~a}}{l(\alpha|\alpha_i) l(\alpha|\alpha_j)} \Phi^a(\alpha_j) 
\label{kinphi2} \end{equation}
where $l(\alpha|\alpha_i)$ is given in eq.(\ref{latticespacing}) and we denote by $g^{ij}(\alpha)$ the metric in $\alpha$ projected in the directions orthogonal to the faces $i$ and $j$:
\begin{equation}
g^{ij}(\alpha) = g^{\mu\nu}(\alpha)~ n_\mu^{(\alpha\cap\alpha_i)}(\alpha)~ n_\nu^{(\alpha\cap\alpha_j)}(\alpha). 
\label{gij} \end{equation}

It should be remarked at this point that  the kinetic term (\ref{kinphi2}), unlike the standard kinetic term of a scalar field on an hypercubic lattice,  involves scalar fields separated by two links, and hence it is quadratic in the gauge variable  $U$. This  occurs also in an hypercubic lattice if a symmetric lattice difference is used as a lattice derivative. 

In the present formulation the two links coupling is required by the choice of the derivative (\ref{derivative3}) and it seems a necessary ingredient to couple scalar fields to the metric.

We consider now the Yang-Mills action coupled to a curved metric $g_{\mu\nu}(x)$:
\begin{equation}
S_{YM} = \int d^d x \sqrt{g(x)}~ {\rm Tr} \left[ F_{\mu\nu}(x) F_{\rho\sigma}(x)\right] g^{\mu\rho}(x) g^{\nu\sigma}(x).
\label{YM} \end{equation} 

In order to put this action on a simplicial lattice we first proceed to construct the lattice analogue of the gauge curvature $F_{\mu\nu}(x)$ following a procedure similar to the one used for the Riemann curvature in Section \ref{Riemannsect}.

Let $\alpha$ be a simplex and $\alpha_i$ the $d+1$ simplices that have with $\alpha$ a face in common. We define, as in Section \ref{Riemannsect}, the hinge $h_{ij}$ as the hinge intersection of $\alpha$,$\alpha_i$ and $\alpha_j$ and the path $\gamma_{h_{ij}}$ as the closed path around $h_{ij}$ starting and ending in $\alpha$. A precise definition is given in (\ref{closedpath}).

We consider now the product of the link variables $U(\alpha|\beta)^a_{~b}$ along the path $\gamma_{h_{ij}}$ and define:
\begin{equation}
U(\gamma_{h_{ij}}|\alpha)^a_{~b} = U(\alpha|\alpha_i)^a_{~c_1}U(\alpha_i|\beta_1)^{c_1}_{~c_2}\dots U(\beta_h|\alpha_j)^{c_{h+1}}_{~c_{h+2}} U(\alpha_j|\alpha)^{c_{h+2}}_{~b}.   \label{Ugamma} \end{equation}

The path $\gamma_{h_{ij}}$ begins and ends in $\alpha$ so that $U(\gamma_{h_{ij}}|\alpha)^a_{~b}$ transforms as follows:
\begin{equation}
U(\gamma_{h_{ij}}|\alpha)^a_{~b} \Rightarrow ( e^{i \eta^A(\alpha) T_A} )^a_{~c}U(\gamma_{h_{ij}}|\alpha)^c_{~f}  ( e^{-i \eta^A(\alpha) T_A} )^f_{~b}  \label{Ugtransf} \end{equation}
so that its trace is invariant under gauge transformations.
Notice also that the orientation of the path is relevant and
\begin{equation}
U(\gamma_{h_{ji}}|\alpha) = U^{-1}(\gamma_{h_{ij}}|\alpha).
\label{Uinv} \end{equation}
 
 Following the same procedure already used for the Riemann curvature tensor we proceed to write the gauge curvature tensor $F_{\mu\nu}(x)$ on the simplicial lattice. First we define the field strength associated to  a single hinge $h_{ij}$ as:
 \begin{equation}
 {\cal F}^a_{~b,\mu\nu}(h_{ij}) = \frac{V(h_{ij})}{v(h_{ij})} U^{(-)}(\gamma_{h_{ij}}|\alpha)^a_{~b}  n_{\mu\nu}^{(ij)}(\alpha) 
 \label{Fmunuhinge} \end{equation}
where
\begin{equation}
U^{(-)}(\gamma_{h_{ij}}|\alpha)^a_{~b} =U(\gamma_{h_{ij}}|\alpha)^a_{~b}-U(\gamma_{h_{ji}}|\alpha)^a_{~b} \label{Uminus} \end{equation}
 and then we define ${\cal F}^a_{~b,\mu\nu}(\alpha)$ by summing, with a suitable weight,  over all the hinges that belong to the simplex $\alpha$:
 \begin{equation}
 {\cal F}^a_{~b,\mu\nu}(\alpha) = \sum_{h_{ij}\in \alpha} \frac{v(h_{ij}|\alpha)}{V(\alpha)}  {\cal F}^a_{~b,\mu\nu}(h_{ij}).
 \label{Fmunu} \end{equation}
 The weights $ \frac{v(h_{ij}|\alpha)}{V(\alpha)}$ are the same used in defining the Riemann curvature in eq.(\ref{curvtens3}) and correspond to the ratio of the support volume in $\alpha$ of $h_{ij}$ and the total volume of $\alpha$.

 Notice finally that  $U^{(-)}(\gamma_{h_{ij}}|\alpha)^a_{~b}$ changes sign when the orientation of the hinge is reversed but that is compensated by the antisymmetry of
 $ n_{\mu\nu}^{(ij)}(\alpha)$ under exchange of $i$ and $j$, so in the end each term in the sum at the r.h.s. of (\ref{Fmunu}) does not depend on the orientation of the hinge.
 
 Given the field strength (\ref{Fmunu}), the Yang-Mills action (\ref{YM}) can be formulated on the simplicial lattice by doing the replacements already  used to write higher order gravity actions in Section \ref{HDBD}:
 \begin{equation}
 S_{YM} \Rightarrow S_{YM}^{({\rm latt})} = \sum_{\alpha} V(\alpha) {\rm Tr} \left[  {\cal F}_{\mu\nu}(\alpha)  {\cal F}_{\rho\sigma}(\alpha) \right] g^{\mu\rho}(\alpha) g^{\nu\sigma}(\alpha).  \label{YMlatt}
 \end{equation}
  It is important to remark that each field strength in (\ref{YMlatt}) contains a plaquette variable, so that the action is a sum of terms involving two plaquettes associated in general to different hinges, and so with different orientations. This is very different from the usual one plaquette action of lattice gauge theories on  flat hypercubic lattices. The coupling of two plaquettes seems to be an essential ingredient if the Yang-Mills action has to be embedded in a curved metric and coupled with gravity.

We conclude this section with some remarks about the topological  invariant  $\theta$ term in $4$ dimensions, namely:
\begin{equation}
S_t = \frac{1}{16 \pi^2}\int d^4 x  {\rm Tr} \left[F_{\mu\nu}(x) F_{\rho\sigma}(x)\right] \epsilon^{\mu\nu\rho\sigma}. \label{topolog} \end{equation}
By following the same correspondence already used for Yang-Mills action we  can at least formally write a simplicial lattice analogue of (\ref{topolog}) as:

\begin{equation}
S_t  \Rightarrow S_t^{({\rm latt})} = \frac{1}{16 \pi^2} \sum_\alpha  \frac{V(\alpha)}{\sqrt{\det g(\alpha)}} \epsilon^{\mu\nu\rho\sigma}~ {\rm Tr}  \left[{\cal F}_{\mu\nu}(\alpha)  {\cal F}_{\rho\sigma}(\alpha)\right].
\label{topologsimp} \end{equation}

The correspondence is purely formal in the sense that we cannot expect the topological nature of the continuum term to be preserved on the lattice, nor there is a guarantee, without further investigation, that it will be recovered in the continuum limit.

However the action has some interesting features which are worth describing. Consider first the following identity, which can be easily verified by using the explicit expression of $V^{(h_{ij})}_{\mu_1 \mu_2}(\alpha)$ given in (\ref{hinge})
\begin{equation}
\epsilon^{\mu_1 \mu_2 \rho_1 \rho_2}~ V^{(h_{ij})}_{\mu_1 \mu_2}(\alpha)~ V^{(h_{kl})}_{\rho_1 \rho_2}(\alpha) = 12 \sqrt{\det g_{\mu\nu}(\alpha)} V(\alpha) \bar{\epsilon}^{~ijkl}
\label{esp} \end{equation}
where  $\bar{\epsilon}^{~ijkl}$ with the indices $i,j,k,l$ running from $1$ to $5$ is completely antisymmetric and further defined by  the relation:
\begin{equation}
\sum_{i=1}^5 \bar{\epsilon}^{~ijkl} = 0   \label{esp2} \end{equation}
and by:
\begin{equation}
 \bar{\epsilon}^{~1234} = \pm 1 \label{1234} \end{equation}
where the sign is determined by the sign of $V(\alpha)$ in (\ref{simplexvolume}) where $d$ has been set to $4$.

By replacing (\ref{esp}) into the action at the r.h.s. of (\ref{topologsimp}) we obtain:
\begin{equation}
 S_t^{({\rm latt})} = \sum_\alpha \sum_{i,j,k,l} \bar{\epsilon}^{~ijkl} {\rm Tr} \left[ U^{(-)}(\gamma_{h_{ij}}|\alpha)  U^{(-)}(\gamma_{h_{kl}}|\alpha) \right] \frac{ v(h_{ij}|\alpha)  v(h_{kl}|\alpha)}{ v(h_{ij})  v(h_{kl})}. \label{tpac} \end{equation}

The action (\ref{tpac}) does not contain the metric $g_{\mu\nu}(\alpha)$ explicitely, but a residual dependence on the metric is present in the weight function $v(h_{ij}|\alpha)$ and $ v(h_{ij})$. 
So if we require complete metric independence, as in the original topological action of the continuum theory, the weight factor at the r.h.s. of (\ref{tpac}) should be modified\footnote{The choice of the wheights $v(h_{ij}|\alpha)$ and $ v(h_{ij})$ has a certain degree of arbitrariness. An different choice, alternative to the one given in Section (\ref{Riemannsect}), is for instance given by $v(h|\alpha) =\frac{2 V(\alpha)}{d(d+1)}$ which satisfy (\ref{alphav}) and (\ref{valpha}) with $v(h) = \frac{2 \sum_{\alpha \in h} V(\alpha)}{d(d+1)}$. With this choice the r.h.s. of (\ref{tpac}) would depend only on the volumes of the simplices and hence only on the determinant of the metric.} 
A possible improvement, in this respect, of the action (\ref{tpac}) is to replace the ratio $\frac{ v(h_{ij}|\alpha) }{ v(h_{ij}) }$ with $\frac{1}{n_{h_{ij}}}$ where $n_{h_{ij}}$ is the number of simplices $\alpha$ that insist on the hinge $h_{ij}$. With this choice the action becomes metric independent and reads:
\begin{equation}
 S_t^{({\rm latt})} = \sum_\alpha \sum_{i,j,k,l} \bar{\epsilon}^{~ijkl} {\rm Tr} \left[ U^{(-)}(\gamma_{h_{ij}}|\alpha)  U^{(-)}(\gamma_{h_{kl}}|\alpha) \right] \frac{ 1}{n_{h_{ij}}n_{h_{kl}}}. \label{tpacmi} \end{equation}

\section{Vierbein and local Lorentz invariance.}
\label{vielbainsect}

Given a simplex $\alpha$ it is always possible to perform a change of  coordinates of the form (\ref{gctalpha})  that transforms the metric $g_{\mu\nu}(\alpha)$ into the flat diagonal metric $\eta_{ab}$ (the flat indices will from now on  be  denoted by the letters $a,b,\dots$  to distinguish them from the "curved" indices $\mu,\nu,\dots$ ).
If we name $\xi^a_i$ the new coordinates of the vertices of $\alpha$ the transformation  (\ref{gctalpha}) now reads:
\begin{equation}
\xi^a_i = \Lambda_\nu^a(\alpha) x^\nu_i + \Lambda^a(\alpha)~~~~~~i\in \alpha \label{gctalpha2} \end{equation}
and according to eq. (\ref{gtransform}) the metric is given in terms of $ \Lambda_\nu^a(\alpha)$ by:
\begin{equation}
g_{\mu\nu}(\alpha) = \Lambda_\mu^a(\alpha)\Lambda_\nu^b(\alpha) \eta_{ab}.  \label{vierbein} \end{equation}

It is clear from (\ref{vierbein}) that $\Lambda_\mu^a(\alpha)$ can be interpreted as a vierbein (or $d$-bein), and that given the metric $g_{\mu\nu}(\alpha)$ the vierbein $\Lambda_\mu^a(\alpha)$ is determined only up to a Lorentz transformation (or rotation in euclidean space-time) acting on the flat index $a$.
Similarly the coordinates $\xi^a_i$ are determined up to a Poincar\'e transformation whose translational part is given by $\Lambda^a(\alpha)$ in (\ref{gctalpha2}). 

We assume that the transformation (\ref{gctalpha2}) is done separately and independently in each simplex\footnote{This means that the  transformation (\ref{gctalpha2}) is not the restriction to the simplex $\alpha$ of a general coordinate transformation (\ref{gct}).}, so that the coordinates $\xi^a_i$ of a vertex $i$ regarded as part of a simplex $\alpha$ are in general different from the coordinates of the same vertex regarded as part of a neighbouring simplex $\beta$. In the following we shall denote these coordinates  $\xi^a_{i}(\alpha)$ to avoid ambiguities.

At the end of this procedure  through a transformation of the form (\ref{gctalpha2}) each simplex  of the simplicial manifold is endowed with  a euclidean reference frame determined  up to a Poincar\'e transformation. Geometrical entities are not affected by the choice of the local frame and the resulting theory will exhibit a local Poincar\'e invariance\footnote{This point of view was first developed in  ref.\cite{Caselle:1989hd} and \cite{Kawamoto:1990hk} and many of the results of the present section can be already found there.}. 

Consider now two neighbouring simplices  $\alpha$ and $\beta$ with a  face in common. The transition from the euclidean reference frame in $\alpha$ to the one in $\beta$ is described by a Poincar\'e transformation, so that the coordinates of a generic point in the reference frame of $\alpha$ and $\beta$ are related by:
\begin{equation}
\xi^a(\beta) = \Omega^a_{\ b}(\beta|\alpha) \xi^b(\alpha) + \Omega^a(\beta|\alpha).
\label{Pionc} \end{equation}

Let $i$ and $j$ be two vertices that belong to both $\alpha$ and $\beta$. Then from (\ref{Pionc}) we have:
\begin{equation}
\left(\xi^a_i(\beta) - \xi^a_j(\beta) \right) -  \Omega^a_{\ b}(\beta|\alpha) \left(  \xi^b_i(\alpha) -  \xi^b_j(\alpha) \right) = 0. \label{xi2} \end{equation}
By replacing in (\ref{xi2}) the euclidean reference frame coordinates $\xi^a_i$ with their value  given in eq.(\ref{gctalpha2}) we obtain the following identity for the vielbeins in $\alpha$ and $\beta$:
\begin{equation}
\left( \Lambda^a_\mu(\beta) -  \Omega^a_{\ b}(\beta|\alpha) \Lambda^b_\mu(\alpha) \right) (x^\mu_i - x^\mu_j) = 0~~~~~~~~~~~~~i,j\in \alpha\cap\beta.  \label{vielid} \end{equation}
Eq.(\ref{vielid}) provides the constraints to which the vielbeins belonging to neighbouring  simplices  have to satisfy  and in fact its square reproduces the analogue constraints (\ref{commonface}) satisfied by $g_{\mu\nu}$.

The rotation matrix $\Omega^a_{\ b}(\beta|\alpha)$ (which is a Lorentz rotation in Minkowski metric) is closely related to the matrix $K^\mu_{\ \nu}(\beta|\alpha)$ that defines the parallel transform. In fact if in eq. (\ref{link}) we replace the curved indices $\mu$ and $\nu$ with flat ones we have:
\begin{equation}
 \Omega^a_{\ b}(\beta|\alpha) = \Lambda^a_{\ \rho}(\beta)K^\rho_{\ \sigma}(\beta|\alpha)\Lambda^{-1\sigma}_{\ \ \ \ b}(\alpha)  \label{lorconn} \end{equation}
 where $\Lambda^{-1\sigma}_{\ \ \ \ b}(\alpha)$ is  the inverse of the vierbein. 
Eq. (\ref{lorconn}) can be written in the form
\begin{equation}
\Lambda^a_{\ \mu}(\beta) =  \Omega^a_{\ b}(\beta|\alpha)\Lambda^b_{\ \rho}(\alpha)  K^\rho_{\ \mu}(\alpha|\beta)
 \label{lorconn2} \end{equation}
which is the analogue in the vielbein formalism of eq.(\ref{gtransp}).
 The Lorentz connection  $\Omega^a_{\ b}(\beta|\alpha)$ transforms under local Lorentz transformations (or rotations in a  euclidean space) as the gauge field in  (\ref{Utransf}), namely;
 \begin{equation}
 \Omega(\beta|\alpha)^a_{~b} \Rightarrow ( e^{i \eta^A(\alpha) T_A} )^a_{~c}\Omega(\beta|\alpha)^c_{~f}  ( e^{-i \eta^A(\alpha_i) T_A} )^f_{~b}
 \label{Otransf} \end{equation}
where $T_A$ are the generators of the group which are supposed here to be in the adjoint representation.

Given a field $\Phi^\mathit{a}(\alpha)$  that transforms under a non trivial representation $\mathit{R}$ of the local Lorentz group its covariant derivative  is of the form (\ref{gaugecovder})  but with the gauge field $U(\beta|\alpha)$ replaced by the Lorentz rotation $\Omega(\beta|\alpha)$ in the representation $\mathit{R}$:
\begin{equation}
\hat{D}_\mu \Phi^\mathit{a}(\alpha) = \frac{1}{2} \sum_{i=1}^{d+1} \Omega_{\mathit{R}}(\alpha|\alpha_i)^\mathit{a}_{~\mathit{b}} \Phi^\mathit{b}(\alpha_i) \frac{V_\mu^{(\alpha\cap\alpha_i)}(\alpha)}{V(\alpha)}.  \label{lorcovder} \end{equation}

Notice however that if the field is also a tensor under general coordinate transformations then a parallel transport has to be done at the same time. For instance the covariant derivative of the vierbein $\Lambda^a_{~\mu}(\alpha)$ is given by:
\begin{equation}
\hat{D}_\mu \Lambda^a_{~\nu}(\alpha) = \frac{1}{2} \sum_{i=1}^{d+1} \Omega(\alpha|\alpha_i)^a_{~b} \Lambda^b_{~\rho}(\alpha_i) K^\rho_{~\nu}(\alpha_i|\alpha) \frac{V_\mu^{(\alpha\cap\alpha_i)}(\alpha)}{V(\alpha)}. 
\label{covderfier} \end{equation}
If we apply  (\ref{lorconn}) and then (\ref{sumarea0}) in  (\ref{covderfier}) we find that the r.h.s. is identically zero, namely that the covariant derivative of the d-bein vanishes as expected:
\begin{equation}
\hat{D}_\mu \Lambda^a_{~\nu}(\alpha) = 0.
\label{covderfier0} \end{equation}

In analogy to what was done in (\ref{Amu}), we can define the Lorentz connection as the gauge field associated to $ \Omega(\alpha|\alpha_i)$:
\begin{equation}
\omega_\mu(\alpha|\alpha_i)^a_{~b} = \left[ \Omega(\alpha|\alpha_i)^a_{~b} - \delta^a_{~b} \right] \frac{V_\mu^{(\alpha\cap\alpha_i)}(\alpha)}{V(\alpha)}. 
\label{omegamu} \end{equation}

It is then easy to write the antisymmetric part of (\ref{covderfier0}) in terms the Lorentz connection, using the fact that the contribution coming from the Christoffel symbol is symmetric and disappears. We have then, in total analogy with the continuum case:
\begin{equation}
\hat{\partial}_{[\mu} \Lambda^a_{~\nu]}(\alpha) - \frac{1}{2} \sum_{i=1}^{d+1}   \omega_{[\mu}(\alpha|\alpha_i)^a_{~b} \Lambda^b_{~\nu]}(\alpha_i) =0
\label{mc} \end{equation}
where the square brackets denote the antisymmetrization in the indices $\mu$ and $\nu$.

The curvature ${\cal R}^a_{~b,\mu\nu}(\gamma_{h_{ij}})$ associated to the the Lorentz connection  $\Omega^a_{\ b}(\beta|\alpha)$ is defined following exactly the same prescriptions (\ref{Ugamma}) and (\ref{Fmunu}) used for gauge theories. 
Given a hinge $h_{ij}$ and the path $\gamma_{h_{ij}}$ defined in (\ref{closedpath}) that goes around $h_{ij}$ starting and ending in the simplex $\alpha$, we can define:
\begin{equation}
\Omega(\gamma_{h_{ij}})^a_{~b} = \Omega(\alpha|\alpha_i)^a_{~c_1}\Omega(\alpha_i|\beta_1)^{c_1}_{~c_2}\dots \Omega(\beta_h|\alpha_j)^{c_{h+1}}_{~c_{h+2}} \Omega(\alpha_j|\alpha)^{~c_{h+2}}_{~b}   \label{Ogamma} \end{equation}
and, in agreement with (\ref{Fmunu}):
\begin{equation} {\cal R}^{a}_{~b,\mu\nu}(\alpha) = \sum_{h_{ij}\in \alpha} \frac{V(h_{ij})}{v(h_{ij})} \left(\Omega(\gamma_{h_{ji}})^a_{~b}-\Omega(\gamma_{h_{ji}})^a_{~b}\right) n_{\mu\nu}^{(ij)}(\alpha). 
 \label{Omunu} \end{equation}
 The curvature $ {\cal R}^{a}_{~b,\mu\nu}(\alpha)$ is directly related to the Riemann curvature tensor. In fact from (\ref{lorconn}) it is easy to show that
 \begin{equation}
 {\cal R}^{ab}_{~~\mu\nu}(\alpha) = \Lambda^a_{~\rho}(\alpha)  \Lambda^b_{~\sigma}(\alpha) {\cal R}^{\rho\sigma}_{~~\mu\nu}(\alpha) \label{RR} \end{equation}
 where curved (resp. flat) indices have been raised with $g^{\mu\nu}(\alpha)$ (resp $\eta^{ab}$).
 The curvature scalar is obviously given by:
 \begin{equation}
  {\cal R}(\alpha) = \Lambda^{-1\mu}_{~~~~a}  \Lambda^{-1\nu}_{~~~~b}  {\cal R}^{ab}_{~~\mu\nu}(\alpha). \label{curvscalf} \end{equation}
  
  As already shown in ref.\cite{Caselle:1989hd} the action (\ref{latticeaction}) can be written in terms of the vielbeins in a form that exhibits a local Lorentz invariance:
  \begin{eqnarray}
&S_l =  \frac{k}{2 (d-2)!} \sum_{\alpha} {\cal R}^{a_1 a_2}_{\mu_1 \mu_2}(\alpha) \Lambda^{a_3}_{~\mu_3}(\alpha)\dots \Lambda^{a_d}_{~\mu_d}(\alpha)~~~~~~~~~~~~~~~~~~~~~~~~~~~~~~~~~~ \nonumber \\ &~~~~~~~~~~~~~~~~~~~~~~~~~~~~ (x_{i_1} - x_{d+1})^{\mu_1}\dots (x_{i_d} - x_{d+1})^{\mu_d} \epsilon_{a_1 a_2 \dots a_d} \epsilon^{i_1 i_2 \dots i_d}.        \label{actionviel} \end{eqnarray}
In showing the equivalence of (\ref{actionviel}) and (\ref{latticeaction}) one uses a trivial consequence of (\ref{vierbein}), namely:
\begin{equation}
\det \Lambda^a_\mu(\alpha) = \sqrt{\det g_{\mu\nu}(\alpha)}. \label{vieldet} \end{equation}
Notice the analogy of (\ref{actionviel}) with the continuum action written in terms of the vielbein and differential form, with the differences $  (x_{i} - x_{d+1})^{\mu_i}$ playing the role of the differentials $d x^\mu$ of the continuum.
 
\section{Coupling of gravity with fermions.}
\label{fermionsect}
The local Lorentz (rotational) symmetry was introduced in the previous section alongside with the vielbein formalism for the metric  by endowing each simplex with an independent  euclidean reference frame.  This allows us to introduce fields that transform as spinors under the local Lorentz transformations, which  is indeed a necessary step if one wants to couple fermionic fields to the metric.

Let $\psi^{\dot{\mathit{a}}}(\alpha)$ be a fermionic field which transforms as a spinor\footnote{Spinorial indices will be denoted with dotted italic letters.} under local $d$-dimensional rotations but is invariant under general coordinate transformations.

Its covariant derivative is then a particular case of eq.(\ref{lorcovder}), namely:
\begin{equation}
\hat{D}_\mu \psi^{\dot{\mathit{a}}}(\alpha) = \frac{1}{2} \sum_{i=1}^{d+1} \Omega_{\mathit{S}}(\alpha|\alpha_i)^{\dot{\mathit{a}}}_{~\dot{\mathit{b}}} \psi^{\dot{\mathit{b}}}(\alpha_i) \frac{V_\mu^{(\alpha\cap\alpha_i)}(\alpha)}{V(\alpha)}  \label{psicovder} \end{equation}
where the label $\mathit{S}$ denotes that the rotation $\Omega(\alpha|\alpha_i)$ is now in a spinorial representation.

The action of a free fermion coupled to the metric is given in the continuum by:
\begin{equation}
S_f = \int d^d x~ \left(\det \Lambda^b_\nu(x) \right)~ \bar{\psi}(x) \gamma^a \Lambda^{-1~\mu}_a(x) D_\mu \psi(x) \label{freeferm} \end{equation}
where $\gamma^a$ are $d$-dimensional $\gamma$ matrices and spinorial indices are understood.

Following the correspondence used already in previous sections for other matter fields we can write (\ref{freeferm}) on the lattice as:
\begin{equation}
S_f^{(\mathrm{latt})} = \sum_{\alpha} V(\alpha)~  \bar{\psi}(\alpha) \gamma^a \Lambda^{-1~\mu}_{a}(\alpha) \hat{D}_\mu \psi (\alpha) \label{freefremlatt} \end{equation}
where the covariant derivative at the r.h.s. is given by (\ref{psicovder}).

The covariant derivative (\ref{psicovder}), and correspondingly the action (\ref{freefremlatt}), can be easily generalized to the case where the fermion transforms also under a representation $\mathit{R}$ of some internal gauge group $G$.
If $U(\alpha,\beta)$ is the gauge field associated this gauge symmetry, then covariant derivative reads:
\begin{equation}
\hat{D}_\mu \psi^{\dot{\mathit{a}}\mathit{r}}(\alpha) = \frac{1}{2} \sum_{i=1}^{d+1} \Omega_{\mathit{S}}(\alpha|\alpha_i)^{\dot{\mathit{a}}}_{~\dot{\mathit{b}}} U_{\mathit{R}}(\alpha|\alpha_i)^{\mathit{r}}_{~\mathit{s}} \psi^{\dot{\mathit{b}}\mathit{s}}(\alpha_i) \frac{V_\mu^{(\alpha\cap\alpha_i)}(\alpha)}{V(\alpha)}  \label{psigaugecovder} \end{equation}
where group elements corresponding to the direct product of $G$ and of the local Lorentz group appear. The action is a direct generalization of (\ref{freefremlatt}) with the covariant derivative (\ref{psicovder}) replaced by (\ref{psigaugecovder}) and the index structure accordingly rearranged.

Another interesting case of fermionic field coupled to gravity is that of a spin $3/2$ field. This  will in fact provide the fermionic (gravitino) term of the $N=1$ four dimensional supergravity. Let $\psi_\mu^{\dot{\mathit{a}}}(\alpha)$ be the spin $3/2$ field on the simplicial lattice. 
It's a covariant vector under general coordinates transformations and transforms as a spinor under local Lorentz transformations. Hence its covariant derivatives is (see also eq.(\ref{covderfier})):
\begin{equation}
\hat{D}_\mu \psi_\nu^{\dot{\mathit{a}}}(\alpha) =  \frac{1}{2} \sum_{i=1}^{d+1} \Omega(\alpha|\alpha_i)^{\dot{\mathit{a}}}_{~\dot{\mathit{a}}} \psi^{\dot{\mathit{b}}}_{~\rho}(\alpha_i) K^\rho_{~\nu}(\alpha_i|\alpha) \frac{V_\mu^{(\alpha\cap\alpha_i)}(\alpha)}{V(\alpha)}. 
\label{covdergravit} \end{equation}

Let us restrict ourselves now to the four dimensional case. The gravitino term of the $N=1$ supergravity action is in the continuum:
\begin{eqnarray}
&S&_{\mathrm{gravit}} = \int d^4x~ \det \Lambda^a_\mu ~ \bar{\psi}_{\mu_1}(x)          \gamma^{a_1 a_2 a_3} \Lambda^{\mu_1}_{a_1}(x) \Lambda^{\mu_2}_{a_2}(x) \Lambda^{\mu_3}_{a_3}(x) D_{\mu_2} \psi_{\mu_3}(x) \nonumber \\ &=&   \int  \bar{\psi}_{\mu_1}(x)~ \epsilon_{a_1 a_2 a_3 a_4}~ \gamma^{a_1 a_2 a_3} \Lambda^{a_4}_{\mu_2}(x) ~D_{\mu_3} \psi_{\mu_4}(x)~ dx^{\mu_1}\wedge dx^{\mu_2}\wedge dx^{\mu_3}\wedge dx^{\mu_4}  \label{gravitaction} \end{eqnarray}
where $\gamma^{a_1 a_2 a_3}$ is the antisymmetrized product of three gamma matrices and $\Lambda^\mu_a$ is the inverse of the vierbain $\Lambda^a_\mu$.
Notice that due to the antisymmetrization of the covariant indices in the covariant derivative the Christoffel symbol does not contribute and the covariant derivative is simply given by:
\begin{equation}
D_{[ \mu} \psi_{\nu ]} = \partial_{[\mu} \psi_{\nu ]}(x) + \omega^{ab}_{[ \mu}(x) \Sigma_{ab}  \psi_{\nu ]}(x).  \label{covp} \end{equation}
where $ \omega^{ab}_{ \mu}(x)$ is the Lorentz connection and $\Sigma_{ab}$ the generators of the Lorentz group in spinorial representation.

The lattice version of the action (\ref{gravitaction}) in the two forms given above can be easily derived by the  usual replacements:
\begin{eqnarray}
&S&_{\mathrm{gravit}}^{\mathrm{latt}} = \sum_\alpha V(\alpha) \bar{\psi}_{\mu_1}(\alpha) \gamma^{a_1 a_2 a_3} \Lambda^{\mu_1}_{a_1}(\alpha) \Lambda^{\mu_2}_{a_2}(\alpha) \Lambda^{\mu_3}_{a_3}(\alpha) \hat{D}_{\mu_2} \psi_{\mu_3}(\alpha) \nonumber \\  &=&  \sum_\alpha \bar{\psi}_{\mu_1}(\alpha)~ \epsilon_{a_1 a_2 a_3 a_4}~ \gamma^{a_1 a_2 a_3} \Lambda^{a_4}_{\mu_2}(\alpha) ~\hat{D}_{\mu_3} \psi_{\mu_4}(\alpha) v^{\mu_1 \mu_2 \mu_3 \mu_4} \label{gravitactionlatt}  \end{eqnarray}
where
\begin{equation}
v^{\mu_1 \mu_2 \mu_3 \mu_4} = \frac{1}{4!} \epsilon^{i_1 i_2 i_3 i_4} (x^{i_1}-x^5)^{\mu_1}(x^{i_2}-x^5)^{\mu_2}(x^{i_3}-x^5)^{\mu_3} (x^{i_4}-x^5)^{\mu_4}.
\label{v} \end{equation}
As in the continuum case the Christoffel symbol in the covariant derivative does not contribute due to the antisymmetrization of the indices, so the expression (\ref{covdergravit}) can be replaced in (\ref{gravitactionlatt}) by:
\begin{equation}
\hat{D}_{[\mu} \psi_{\nu]}^{\dot{\mathit{a}}}(\alpha) =  \frac{1}{2} \sum_{i=1}^{d+1} \Omega(\alpha|\alpha_i)^{\dot{\mathit{a}}}_{~\dot{\mathit{a}}} \psi^{\dot{\mathit{b}}}_{~[\nu}(\alpha_i) \frac{V_{\mu]}^{(\alpha\cap\alpha_i)}(\alpha)}{V(\alpha)}. 
\label{covdergravitanti} \end{equation}

By adding the action for pure gravity given in (\ref{actionviel}) to the gravitino action as given above in (\ref{gravitactionlatt}) one can write a lattice action that corresponds in the continuum to the $N=1$ supergravity in $4$ dimensions:
\begin{equation}
S_{\mathrm{sugra}} = \frac{k}{4} \sum_{\alpha} \epsilon_{a_1 a_2 a_3 a_4} \left[  {\cal R}^{a_1 a_2}_{\mu_1 \mu_2}(\alpha) \Lambda^{a_3}_{~\mu_3}(\alpha) \Lambda^{a_4}_{~\mu_4}(\alpha) + \gamma^{a_1 a_2 a_3} \Lambda^{a_4}_{\mu_2}(\alpha) ~\hat{D}_{\mu_3} \psi_{\mu_4}(\alpha) \right]  v^{\mu_1 \mu_2 \mu_3 \mu_4} \label{sugra4_1} \end{equation}
where $v^{\mu_1 \mu_2 \mu_3 \mu_4}$ is given in (\ref{v}).
It is important to remark that although the action (\ref{sugra4_1}) is formally analogue in the present formalism to the continuum $N=1$ four dimensional supergravity, exact supersymmetry is certainly broken\footnote{For instance it is crucial in the continuum that the commutator of two covarint derivatives is proportional to the curvature, which is not true here, since a whole loop around a hinge is needed to reproduce the curvature.} on the lattice and there is no guarantee at this stage that it would be recovered in the continuum limit. This should be the object of an independent investigation.

\section{Coupling of gravity to differential $p$-forms}
\label{pformsection}
In the previous sections we described the coupling of different types of matter fields (scalar fields, gauge fields, fermions) with gravity within the simplicial lattice framework of the Regge calculus. This has provided us with a dictionary to translate any continuum action containing those fields into a simplicial lattice action.

For the correspondence to be complete however it would still be necessary to find a lattice description  of fields that are differential $p$-forms (with $p > 1$).  These fields play an important role in many relevant theories, for instance a 3-form field $A_{[\mu\nu\rho]}(x)$ is one of the fundamental fields of supergravity in $11$ dimensions.

In general a differential $p$-form field $A_{[ \mu_1 \mu_2 \dots \mu_p]}(x)$ is associated to an abelian gauge invariance of the form:
\begin{equation}
 A_{[ \mu_1 \mu_2 \dots \mu_p]}(x) \Longrightarrow A_{[ \mu_1 \mu_2 \dots \mu_p]}(x) + \partial_{[ \mu_1} \Lambda_{\mu_2 \dots \mu_p]}(x)  \label{pformtr} \end{equation}
 where the gauge parameter $\Lambda_{[\mu_1 \dots \mu_{p-1}]}(x)$ is a $p-1$ form.
 
 The gauge invariant field strength is then a $p+1$ form and is given by:
 \begin{equation}
 F_{[\mu_1 \mu_2 \dots \mu_{p+1}]}(x) = \partial_{[ \mu_1} A_{ \mu_2 \mu_3 \dots \mu_{p+1}]}(x)  \label{pformstrength} \end{equation}
where the square brackets denote antisymmetrization of the indices. 

The most direct  way to write a $p$ form on a simplicial lattice following the approach described in the previous sections would be to replace the continuum field $ A_{[ \mu_1 \mu_2 \dots \mu_p]}(x)$ with a completely antisymmetric tensor $ A_{[ \mu_1 \mu_2 \dots \mu_p]}(\alpha)$ of rank $p$ associated to each simplex $\alpha$, and to define its field strength as its covariant derivative, which in this case would  coincide with the ordinary derivative due to the antisymmetrization of the indices. In short, eq.(\ref{pformstrength}) would be replaced by:
\begin{equation}
 F_{[\mu_1 \mu_2 \dots \mu_{p+1}]}(\alpha) = \hat{\partial}_{[ \mu_1} A_{ \mu_2 \mu_3 \dots \mu_{p+1}]}(\alpha).  \label{pformstrength2} \end{equation}

 However with this definition a gauge transformation of the type (\ref{pformtr}), with $x$ replaced by $\alpha$ and the partial derivative by the lattice derivative (\ref{derivative3}), would not be a symmetry of the field strength.
 In fact one can easily see from the definition of the partial derivative on a simplicial lattice given in (\ref{derivative3}) that derivatives in different directions do not commute, namely:
\begin{equation}
\hat{\partial}_\mu \left(\hat{\partial}_\nu \phi(\alpha) \right) \neq \hat{\partial}_\nu \left(\hat{\partial}_\mu \phi(\alpha) \right).  \label{nocomm} \end{equation}
 
 This is a consequence of the simplicial lattice structure: derivatives are associated to  one link moves on the dual lattice, which is made of Voronoi cells, and on such lattice the result of two moves depends on their order, unlike what happens on a hypercubic lattice.

In the previous sections the one forms describing gauge fields have been associated to the $(d-1)$-dimensional faces of the  simplices, that is to the links of the dual Voronoi tasselation.
Similarly the two form describing curvatures or field strengths  were associated to the $d-2$ dimensional hinges, namely to the two dimensional plaquettes of the dual lattice.

It is clear then that the natural way to describe a $p$-form field on a simplicial lattice would be to associate it to a $p$-dimensional cell of the dual Voronoi tasselation. This is completely identified by the $d-p+1$ vertices of its dual $d-p$ dimensional simplex\footnote{For a more precise definition of this simplex-cell duality see for instance ref. \cite{Dec} }.

The problem is then to formulate on the simplicial lattice a discrete exterior calculus, endowed with a wedge product of forms and of a nilpotent differential $d$ operator that satisfy, as much as possible, the usual algebraic properties of the exterior calculus.

This problem has been investigated (see for instance \cite{Dec} and references therein) but mostly in the more direct way of associating a $p$-form to a $p$-dimensional simplex of the simplicial complex.

It was shown in ref.\cite{Dec} that  a wedge product of a $p$ and a $q$ form can be defined as a quantity associated to $p+q$ dimensional simplices. This product is commutative (in a graded sense) as in the continuum, but it is not associative, although the non-associative terms can be shown to vanish in the continuum limit.  Finally a $d$ operator can be defined, that satisfies the nilpotency relation $d^2=0$ and the graded distributive property with respect to the wedge product.

However the case we are interested in is different:  a $p$-form has  now to be associated to a $p$ dimensional Voronoi cell, which is dual to a $d-p$ dimensional simplex within the simplicial complex . 
A $p$-form is then a field defined on the $d-p$ dimensional simplices, and the wedge product  of a $p$-form and a $q$-form  should be associated to  $d-p-q$ dimensional simplices, which are dual to $p+q$ dimensional Voronoi cells. 

Vertices in a Voronoi cell, which correspond to $d$-dimensional simplices on the original lattice, can be several links apart, unlike what happens on a simplex where all pairs of vertices are connected by a link.  This can make defining  a wedge product  and a differential $d$  operator that satisfy the algebraic rules of exterior calculus even more difficult than in the case where the forms are defined directly on simplices.

This is indeed the case.
We succeeded in defining a wedge product for forms (see the Appendix for details) defined on Voronoi cells and also a nilpotent differential operator $\hat{d}$ that maps a $p$ form into a $p+1$ form.

 This  ensures that by operating with $\hat{d}$ on a given $p$ form gauge field one obtains a $p+1$ form - the field strength - which is a  invariant under gauge transformations whose parameters are $p-1$ forms thus overcoming the problem discussed at the beginning of this section.

However the wedge product defined in this way has some rather severe shortcomings. For a start, as in the case mentioned above of the wedge product of forms defined directly on the simplices, it is not associative.   More worryingly the differential operator $\hat{d}$ does not satisfy the Leibnitz rule when applied to the wedge product of forms. 

This implies that, although gauge invariance is preserved, partial integration is not allowed\footnote{The Leibnitz rule, and hence partial integration, might be recovered in the continuum limit, but further investigation is needed in that respect.} and different forms of an action, which are equivalent in the continuum up to surface terms may become different on the lattice.

This is particularly important in actions like the Chern-simons action in three dimensions or the $\int FFA$ term (with $A$ the above mentioned three form and $F$ its field strength) in eleven dimensional supergravity.  In the continuum these actions can be written as surface terms of gauge invariant actions in  respectively three and twelve dimensions, but this property breaks down if the Leibnitz rule is violated.

In spite of its shortcomings the above mentioned wedge product is interesting and may be the base for future investigations in the subject, particularly concerning the recovery of the fundamental algebraic properties of the exterior calculus in the continuum limit.
For this reason the details of its definition and of its main properties are given in the Appendix.

\section{Some final remarks.}
\label{conclusions}

This paper started as an attempt to answer a perhaps naive question: "Is it possible to have a formulation of simplicial gravity where the fundamental degrees of freedom are, as in the continuum theory, the components of the metric tensor?". 

Since the metric tensor must depend on the choice of coordinates, we had to attach coordinates  to the vertices of the simplices and require  invariance under coordinates transformations. This discrete version of invariance under coordinate transformations does not imply invariance under diffeomorphisms, as the vertices form a discrete set, their adjacency matrix is kept fixed and the model is ultimately equivalent to Regge Calculus. However the invariance under coordinate transformations provides the basis for a discrete tensor calculus which, in turns, makes the correspondence with the continuum theory much more strict and suitable for extension to the coupling of gravity to different types of matter fields.

One crucial ingredient of this correspondence is the definition of partial derivative on the lattice defined in Section \ref{derivativessl}. This  can be regarded as a generalization to simplicial lattices of the symmetric finite difference operation on an hypercubic lattice and is strongly motivated by the requirement that it transforms as a covariant vector under general coordinate transformation. 

The original aim turns then into a more ambitious one, namely finding a precise correspondence, a kind of dictionary, between actions in the continuum and actions on a simplicial lattice, thus allowing to write the coupling of any matter field to gravity within the framework of Regge Calculus. Following this correspondence we were for instance able to write an action on the simplicial lattice that correponds in the continuum to supergravity in $4$ dimensions.

The problem of coupling scalars, fermions and gauge fields to discrete gravity has obviously been discussed in the literature before (see references in the different sections) but mostly on a case by case basis, without a unifying scheme as the one we developed here. However, as discussed in the last section and in the Appendix , the correspondence with the continuum is not complete: actions that contain $p$-form potentials gauging free differential algebras do not seem to fit in this scheme and coupling them consistently to gravity within the framework of Regge calculus is still an open problem. 

Much work still needs to be done. We have not checked for instance the continuum limit, even at the classical level, of the actions of the different kinds of matter coupled to gravity. This is particularly relevant for gauge theories. In fact the simplicial lattice action for pure Yang-Mills theory is quite different, even in absence of gravity, from the traditional Wilson action as it consists of two plaquette terms rather than of the usual one plaquette term.

Although the correspondence with the continuum theory is quite compelling there are some fundamental differences that would also need further investigation. The fundamental degrees of freedom in our approach are the component of the metric tensor on each simplex, but these are not independent degrees of freedom as they are constrained to coincide on their common  $d-1$ dimensional faces. As a result the $d(d+1)/2$ degrees of freedom of the components of $g_{\mu\nu}(x)$ at the point of coordinates $x^\mu$ are spread on the simplicial lattice over a number of neighbouring simplices which is of order $d$. This would  obviously be relevant in any attempt to find a correct measure of integration in a functional integral for quantum gravity. We have not addressed this problem here. 

A lattice length $l(\alpha|\beta)$ has been defined in (\ref{latticespacing}) and some symmetries of the continuum theory are broken by higher order terms in $l(\alpha|\beta)$ and are recovered in the limit where $l(\alpha|\beta)$ tends to zero. As already remarked this is the case of some symmetries of the Riemann tensor which are violated on the lattice by higher order terms in the deficit angle $\theta_h$. The presence of higher order terms makes it also  apparently impossible to invert eq.(\ref{qq}) and express the Christoffel symbol in terms of derivatives of the metric tensor.

\vspace{1cm}

{\bf{\Large Acknowledgments}}

\vspace{0.5cm}
I wish to thank  M. Billo,  M. Caselle and  N. Kawamoto for discussions and critical reading of the manuscript.
\vspace{1cm}\vspace{1cm}
\begin{center}
{\bf{\Large Appendix}}
\end{center}

\appendix
\section{An attempt of constructing a discrete exterior calculus within the Regge Calculus framework.}
\label{appendix}

 In describing the interaction of matter fields with gravity within the framework of Regge Calculus scalar fields (zero forms) have been associated to the $d$-dimensional simplices and the gauge fields (one forms) to their $(d-1)$ dimensional faces, namely they have been respectively associated to the sites and the links of the dual Voronoi lattice.
 The Voronoi tassellation generated by the vertices of the simplicial lattice consists of $d$-dimensional cells which are dual to the vertices. The $p$-dimensional faces of the Voronoi cells are dual to the $d-p$ dimensional simplices of the original lattice, and each of them is completely identified 
by the $d-p+1$ vertices of the dual simplex. 
The natural generalization of the $p=0$ and $p=1$ cases to arbitrary $p$ is to associate a $p$-form field of the continuum theory to the $p$-dimensional cells of the dual Voronoi tassellation, namely, by duality, to the $d-p$ dimensional simplices of the original simplicial lattice. More precisely if we denote by $\sigma_{d-p}$  the $d-p$ dimensional simplices and by $\star\sigma_{d-p}$ the $p$ dimensional cells dual to them, we can define a discrete $p$-form as a map from $\star \sigma_{d-p}$ onto the real numbers.

 The simplex $\sigma_{d-p}$ is identified by its $d-p+1$ vertices $P_0, P_1, \dots P_{d-p}$:
 \begin{equation}
 \sigma_{d-p} \equiv [P_0, P_1, \dots P_{d-p}].  \label{psimplex} \end{equation}
 Similarly we shall identify $\star \sigma_{d-p}$ as:
 \begin{equation}
\star \sigma_{d-p} \equiv \star [P_0, P_1, \dots P_{d-p}].  \label{pcell} \end{equation}
 A $p$-form field $A_{[\mu_1 \mu_2 \dots \mu_p]}(x)$ of the continuum theory will have the following correspondence on the simplicial lattice:
 \begin{equation}
 A_{[\mu_1 \mu_2 \dots \mu_p]}(x) \Longrightarrow  A(\star[P_0, P_1, \dots P_{d-p}]).
 \label{pform} \end{equation}
 Notice that the simplex $ \sigma_{d-p}$ and the cell $\star \sigma_{d-p}$ are oriented, so the map defined by eq.(\ref{pform}) is antisymmetric under permutations of the vertices, for instance:
 \begin{equation}
  A(\star[P_0, P_1, P_2 \dots P_{d-p}]) = -  A(\star[P_1, P_0, P_2 \dots P_{d-p}]).
  \label{antiform} \end{equation}
  
  In order to procede with the construction of the discrete theory we need to set up and define at least the basic ingredients of the discrete exterior calculus\footnote{The discrete exterior calculus that we try to construct here is different from the one extensively discussed for instance in ref.\cite{Dec}, since we associate $p$-forms to $d-p$ dimensional simplices (or $p$ dimensional Voronoi cells) rather than to $p$ dimensional simplices as in ref.\cite{Dec}. }.

Let us first introduce the notion of discrete exterior derivate. The exterior derivative of a $p$-form is a $p+1$ form, hence it is defined on the $p+1$ dimensional cells of the dual Voronoi tassellation or equivalently by duality on the $d-p-1$ dimensional simplices of the original lattice.

Given the $p$-form $A$ at the r.h.s. of (\ref{pform}) its exterior derivative $\hat{d}A$ is then a function of the ordered $d-p$ vertices of a $d-p-1$ dimensional simplex, and it can be defined as:
\begin{equation}
\hat{d}A(\star[P_1,P_2,\dots,P_{d-p}]) = \sum_{Q} A(\star[P_1,P_2,\dots,P_{d-p},Q]) \label{extder} \end{equation}
where the sum is extended to all vertices $Q$ such that $[P_1,P_2,\dots,P_{d-p},Q]$ is a simplex that has $[P_1,P_2,\dots,P_{d-p}]$ as a proper face. In terms of the dual lattice the sum at the r.h.s of (\ref{extder}) is over the $p-1$ dimensional cells $\star[P_1,P_2,\dots,P_{d-p},Q]$ that form the boundary of the $p$ dimensional cell $\star[P_1,P_2,\dots,P_{d-p}]$.

Eq.(\ref{extder}) can be generalized to the $p$-chains $\omega_p$ defined as finite formal sums of the $p$-cells with coefficients in ${\mathbb Z}$:
\begin{equation}
\omega_p = \sum_i l_i~\star \sigma^i_{d-p} = \sum_i l_i~  \star[P^i_0, P^i_1, P^i_2 \dots P^i_{d-p}]. \label{pchain} \end{equation}

The boundary operator $\partial$ acts on $\omega_p$ as:
\begin{equation}
\partial \omega_p =  \sum_i l_i~ \sum_{Q^i} \star[P^i_1,P^i_2,\dots,P^i_{d-p},Q^i]
\label{boundary} \end{equation}

Assuming that the map defining the $p$-form $A$ is a linear one, we can generalize (\ref{extder}) to the form:
\begin{equation}
\hat{d}A(\omega_p) = A(\partial \omega_p) \label{stokes} \end{equation}
which is the discrete equivalent of
\begin{equation}
\int_{{\mathcal M}_p} dA_p = \int_{\partial{\mathcal M}_p} A_p. \label{stokes2} \end{equation}

From the definition (\ref{boundary}) and the antisymmetry (\ref{antiform}) it follows immediately that the square of the boundary operator is zero, and consequently that also $\hat{d}^2=0$.

The next step is define a wedge product of two discrete forms trying to preserve  as much as possible the algebraic properties of the product of forms in the continuum. 
The product of a $p$-form and a $q$-form is a $(p+q)$-form, so in our discrete formalism it should be of the form:
\begin{equation}
A(\star\sigma_{d-p}) \wedge B(\star\sigma_{d-q}) \Longrightarrow (A\wedge B)(\star\sigma_{d-p-q}). \label{wedge1} \end{equation}
The best definition of discrete wedge product we could find has the form\footnote{This wedge product is also considered in \cite{Dec} as the "discrete dual-dual wedge product", but its properties are not studied there.}:
\begin{eqnarray}
&(A\wedge B)(\star[P_0,P_1,\dots,P_{d-p-q}]) = \sum_{R_1,\dots,R_p,S_1,\dots,R_q} A(\star[P_0,\dots,P_{d-p-q},S_1,\dots,S_{q}]) \nonumber \\ &B(\star[P_0,\dots,P_{d-p-q},R_1,\dots,R_{p}])~ E([P_0,\dots,P_{d-p-q},S_1,\dots,S_{q},R_1,\dots,R_{p}])  \label{wedge2} \end{eqnarray}
where $E([P_1,P_2,\dots,P_{d+1}])= \pm 1$ if the $d+1$ vertices $P_1,P_2,\dots,P_{d+1}$ form a $d$-dimensional simplex, otherwise it is zero.

The symbol $E([P_1,P_2,\dots,P_{d+1}])$ is completely antisymmetric in its arguments and the $\pm$ sign may be chosen to coincide with the sign of the volume in eq.(\ref{simplexvolume}). While an overall sign in the definition of $E([P_1,P_2,\dots,P_{d+1}])$ is essentially a matter of convention, the relative sign between two neighbouring simplices  is crucial and is given by:
\begin{equation}
E([P_1,P_2,\dots,P_i,\dots,P_{d+1}]) = - E([P_1,P_2,\dots,P'_i,\dots,P_{d+1}]) \label{relativesign} \end{equation}
where $P_i$ and $P'_i$ are the vertices which are not shared by the two simplices, which have a $d-1$ dimensional face in common.
Repeated use of (\ref{relativesign}) determines in principle the signs of the $E$ simbol for all simplices of the simplicial complex (assuming it is simply connected).

 It follows immediately from (\ref{wedge2}) that even and odd forms (anti)commute according to the usual rule:
\begin{equation}
A(\star\sigma_{d-p}) \wedge B(\star\sigma_{d-q}) = (-1)^{pq} ~ B(\star\sigma_{d-q}) \wedge A(\star\sigma_{d-p}). \label{anticomm} \end{equation}

However some important properties of the wedge product in the continuum are not preserved by (\ref{wedge2}). First of all the product defined in (\ref{wedge2}) is not associative. This was to be expected: the wedge product introduced in \cite{Dec}, where $p$-forms are directly associated to $\sigma_{p}$ rather than to $\star\sigma_{d-p}$ as in our case, was shown not to be associative, although it was proved in the same paper that associativity is recovered in the continuum limit.

In order to show the non associativity of (\ref{wedge2}) it is enough to write explicitely the product of the forms $A(\star\sigma_{d-p})$, $B(\star\sigma_{d-q})$ and $C(\star\sigma_{d-r})$:
\begin{eqnarray}
\left( (A\wedge B)\wedge C \right)(\star[T_0,T_1\dots T_{d-p-q-r}])=&\nonumber \\ \sum_{P_{\dots},Q_{\dots},R_{\dots},S_{\dots}} A(\star[T_{\dots},R_{\dots},Q_{\dots}])\cdot& B(\star[T_{\dots},R_{\dots},P_{\dots}])\cdot 
C(\star[T_{\dots},S{\dots }])\cdot \nonumber \\ E([T_{\dots},R_{\dots},Q_{\dots},P_{\dots}])\cdot E([T_{\dots},R_{\dots} ,S_{\dots}])& \label{noncomm} \end{eqnarray}
where $T_{\dots}$ stands for the set of points $T_0,T_1\dots T_{d-p-q-r}$. Similarly $R_{\dots}$,$Q_{\dots}$,$P_{\dots}$ and $S_{\dots}$ stand for sets of respectively $r$, $q$, $p$ and $p+q$ points.

The non commutativity is apparent from the asymmetry of (\ref{noncomm}) in the three forms $A$, $B$ and $C$. The symmetry would be restored if the set of points $\{ S_{\dots} \}$ coincided with the union of $\{ P_{\dots} \}$ and $\{ Q_{\dots} \}$, so the associative terms correspond to a subset of the terms appearing at the r.h.s. of (\ref{noncomm}). We do not have any argument at the moment to argue that associativity would be restored in the continuum limit, further investigation is needed in that respect.

The other property which is not satisfied by the wedge product (\ref{noncomm}) is the distributive law (Leibnitz rule) with respect to the exterior derivative defined in (\ref{extder}).

As in the case of the non associativity this can be checked directly. Let us consider the wedge product of a $p$ and a $q$ form defined in (\ref{wedge2}) and take its exterior derivative. We have:
\begin{equation}
\hat{d}(A\wedge B)\left([T_{\dots}]\right) \sum_{P_{\dots},Q_{\dots},R} A([T_{\dots},R,Q_{\dots}])\cdot B(T_{\dots},R,P_{\dots}])\cdot E([T_{\dots},R,Q_{\dots},P_{\dots}]) \label{dawb} \end{equation}
where $T_{\dots}$ has now only $d-p-q$ entries, as $\hat{d}(A\wedge B)$ is a $p+q+1$ form, while $P_{\dots}$ and $Q_{\dots}$ are defined as above. The sum over the single vertex $R$ is the result of the exterior derivative operation.

We shall compare the result of (\ref{dawb}) with what one would expect if the Leibnitz rule were valid, namely:
\begin{eqnarray} 
&\left(\hat{d}A\wedge B +(-1)^{p} A\wedge \hat{d}B \right)([T_{\dots}])=\sum_{P_{\dots},Q_{\dots},R,S} A([T_{\dots},R,Q_{\dots}])\cdot \nonumber \\& B([T_{\dots},S,P_{\dots}])\cdot \left\{E([T_{\dots},R,Q_{\dots},P_{\dots}])+E([T_{\dots},S,Q_{\dots},P_{\dots}])\right\}. \label{dadb} \end{eqnarray}

If we compare the r.h.s. of eq.(\ref{dadb}) with the r.h.s. of (\ref{dawb}) we see that in the former there is an extra sum over the vertex $S$ that was not present in (\ref{dawb}). The two expressions have the same structure only in a subset of terms, namely if in (\ref{dadb}) we set $R=S$. 
In fact, while at the r.h.s. of (\ref{dawb}) the forms $A$ and $B$ take  value on simplices that are both contained in the same $d$ dimensional simplex (i.e. the argument of the $E$ function) this is not generally true in eq.(\ref{dadb}) unless  $S$ and $R$ are set to be equal.

The lack of associativity is not a problem in three dimensional Chern Simons theory and in $11$ dimensional supergravity. For instance if $A$ is the three form field of supergravity in $11$ dimension it is immediate to see that even without assuming associativity the two forms $(A\wedge dA)\wedge dA$ and $A\wedge (dA\wedge dA)$ only differ for a total differential, provided the distributive law with respect to $d$ is satisfied.

The violation, by a large number of terms, of the Leibnitz rule is a much more serious problem because it prevents from using partial integration and from writing the Chern Simons action and the $FFA$ term in $11$ dimensional supergravity as boundary terms of topological actions in one higher dimension.

In the descrete exterior calculus described in \cite{Dec} $p$ forms are associated to $p$ dimensional simplices rather than to the $p$ dimensional Voronoi cells of the dual lattice.  The wedge product defined there is not associative but satisfies the distribution law (Leibnitz rule) with respect to the exterior derivative. 

However defining $p$ forms on the $p$ dimensional simplices does not seem to fit in the Regge Calculus scheme outlined in this paper. At the root of the difficulty,  which seems of difficult solution, is the asymmetry between simplicial lattice and dual lattice, which does not allow a consistent definition of a discrete dual Hodge operator.

%

\end{document}